\documentclass[showpacs,preprintnumbers,amsmath,amssymb,superscriptaddress]{revtex4}

\usepackage{color,epsfig} 
\usepackage{graphicx}

\usepackage{rotating}

\def\be{\begin{equation}}
\def\ee{\end{equation}}

\begin{document}
\preprint{\tt MAN/HEP/2006/21}

\title{Constraints on Supersymmetric Hybrid Inflation Models}
\author{Richard A. Battye}
\affiliation{Jodrell Bank Observatory, University of Manchester, Macclesfield, Cheshire SK11
9DL, UK.}
\author{Bj\"orn Garbrecht}
\affiliation{School of Physics and Astronomy, University of Manchester, Oxford Road, Manchester M13 9PL, UK.}

\author{Adam Moss}
\affiliation{Jodrell Bank Observatory, University of Manchester, Macclesfield, Cheshire SK11
9DL, UK.}

\date{\today}
 
\begin{abstract}
We point out that the inclusion of a string component contributing around 5\%
to the CMB power spectrum amplitude on large scales can increase the preferred
value of the spectral index $n_{s}$ of density fluctuations measured by CMB
experiments. While this finding applies to any cosmological scenario involving
strings, we consider in particular models of supersymmetric hybrid inflation,
which predict $n_{s} \stackrel{>}{{}_\sim} 0.98$, in tension with
the CMB data when strings are not included. Using MCMC
analysis we constrain the parameter space allowed for $F$-
and $D$-term inflation.
For the $F$-term model, using minimal supergravity corrections, we find that
$\log\kappa= -2.34\pm 0.38$ and $M= (0.518\pm 0.059)\times 10^{16}{\rm GeV}$.
The inclusion of non-minimal supergravity corrections can modify these values somewhat. In the corresponding analysis
for $D$-term inflation, we find $\log\kappa= -4.24\pm 0.19$ and $m_{\rm FI}= (0.245\pm 0.031)\times 10^{16}{\rm GeV} $. Under the assumption that these models are correct, these results represent precision measurements of important parameters of a Grand Unified Theory.
We consider the possible uncertainties in our measurements and additional constraints on the scenario from the stochastic background of gravitational waves produced by the strings. The best-fitting model predicts a $B$-mode polarization signal $\approx 0.3 \mu {\rm K}$ rms peaking at $\ell\approx 1000$. This is of comparable amplitude to the expected signal due to gravitational lensing of the adiabatic $E$-mode signal on these
scales.
\end{abstract}


\maketitle

\section{Introduction}

The publication of the most recent results from the Wilkinson Microwave Anisotropy Probe~\cite{wmap} (WMAP) has focused attention in the direction of precision constraints on inflationary models~\cite{PrecCon} believed to be the origin of the initial spectrum of density fluctuations~\cite{DensityPerturbations}. By measuring the power spectrum of the $E$-mode polarization of the cosmic microwave background (CMB) on large-scales, the WMAP data constrains the optical depth to reionization, $\tau_{\rm R}=0.09\pm 0.03$, with consequent improvement to parameters such as the spectral index of density fluctuations, $n_{s}=0.951^{+0.015}_{-0.019}$, which are degenerate with $\tau_{\rm R}$. Although ad-hoc inflationary models can be constructed with a range of values of $n_{s}$, some specific models predict very narrow ranges for $n_{s}$ making them vulnerable to the very tight constraints now available.

One such class of models is supersymmetric (SUSY)
hybrid inflation~\cite{hybrid}, which has 
$F$-term~\cite{CLLSW,DSS} and $D$-term~\cite{DTerminf} variants.
These scenarios are particularly attractive since the
potential for the inflaton is flat at tree level. It acquires corrections from
loop effects and when further tree-level, non-renormalisable Planck scale
suppressed operators are added, and it can naturally meet the slow-roll
conditions. Within certain minimal realisations, one is left with the choice
of only one dimensionless coupling constant $\kappa$ and a mass scale $M$. One can deduce the amplitude of curvature perturbations $P_{\cal R}$, their spectral index $n_{s}$ and the dimensionless string tension $G\mu$ from $\kappa$ and $M$ (up to a weak dependence on the unknown reheat temperature, $T_{R}$).
Therefore, these models may be considered as rather predictive when compared
to other scenarios for inflation. 

One can make an analytic estimate of the scalar spectral index, which yields
$n_{s} \stackrel{>}{{}_\sim} 1-{1/N_{\rm e}}$, where $N_{\rm e}$ is the number of e-foldings (measured from the end of inflation) when a particular observed scale leaves the horizon (or conversely when it comes back inside the horizon). For standard estimates of the reheat temperature, $N_{\rm e}\approx 50$, making $n_{s} \stackrel{>}{{}_\sim} 0.98$ a prediction of the simplest version of these models. If one takes the quoted observational constraint seriously, such a model would be excluded at around the $2-\sigma$ level under the assumption that only adiabatic density fluctuations are created during inflation.
However, in these models cosmic strings will be formed at the end of inflation,
if the phase transition induces the spontaneous breakdown of a ${\rm U}(1)$ symmetry
by the waterfall fields. Since the predicted energy scale of these strings is
around the grand unified (GUT) scale, they may contribute significantly to
the observed perturbations~\cite{jean97}, creating an interesting phenomenology~\cite{WB,CHMb}.

In this paper, we point out that the inclusion of a sub-dominant string contribution of around 5\% to the large scale power spectrum amplitude of the CMB can
increase the preferred value for the spectral index up to $n_s\approx 0.98$
(and the maximum allowed value at $2-\sigma$ level up to $n_s\approx 1.02$),
something which is a generic point valid for all models of inflation which
produce cosmic strings. Naively, it may seem to be a grotesque violation of Occam's razor to have two sources of fluctuations with nearly equal amplitude; in no way do we claim that the data requires the additional string contribution in a Bayesian model selection sense. But as we shall describe,
a string contribution of the required size arises very naturally in the class of models under consideration here.  Note moreover, that these models constitute an attempt of fusing together the areas of particle physics and cosmology and may therefore be considered to be more attractive when put in the wider context.
We then proceed to constrain the parameters $\kappa$ and $M$ for specific realizations of $F$-term and $D$-term hybrid inflation models. Our results show that the inclusion of the string contribution is critical to determining the correct constraints on the parameters.

We note that the upper bound on the string tension $\mu$ has recently been discussed by a number of authors~\cite{pog,fraisse05,fraisse06,slosel,bevis}. Their basic conclusion, using a variety of different methods, has been that there is a $2-\sigma$ upper limit of $G\mu < (2-3)\times 10^{-7}$, something which we shall confirm. However, they have ignored the effect strings have on the preferred value of  $n_{s}$.
Qualitatively similar ideas were pointed out in ref.~\cite{hind} in the context of constraints on the global texture model using the first year WMAP data. At that point in time, the constraint on $n_{s}$ was not as tight as is the case now and, therefore, the necessity of including the defect contribution was not so critical.

\section{Models and methodology}

In Section~\ref{Fterm} and~\ref{Dterm} we introduce the inflationary models
we are studying and discuss in particular the various contributions to the
inflaton potential. In~\ref{Sspec} we specify the model for the string
power spectrum we are using
and in~\ref{MCMC} the details of the 
Markov-Chain-Monte-Carlo (MCMC) analysis are presented.

\subsection{$F$-term inflation}
\label{Fterm}

$F$-term inflation~\cite{CLLSW,DSS} is implemented by the superpotential
\begin{equation}
W=\kappa \widehat S ({\widehat{\overline G}} \widehat G -M^2)\,,
\end{equation}
where $\widehat S$ denotes a gauge-singlet chiral superfield,
$\widehat G$  belongs to a certain $\cal N$-dimensional representation of
the gauge group and ${\overline{\widehat G}}$ is a corresponding conjugate
multiplet. In the subsequent discussion, we set ${\cal N}=1$, as we are
studying the gauge group ${\rm U}(1)$, the breaking of which leads to the
production of cosmic strings.

Having specified the model, we now enumerate the various contributions to
the scalar potential, which determine the inflationary dynamics.
The leading order contribution is the tree-level scalar potential
\begin{equation}
V_0= \kappa^2 \bigg[\left| \overline G G - M^2\right|^2
+ \left|S \overline G \right|^2 + \left|S G\right|^2\bigg]\,,
\end{equation}
where $G$, $\overline{G}$ and $S$ are the scalar components of the respective
superfields. While $S$ is to be identified with the inflaton, $G$ and
$\overline{G}$ are usually referred to as waterfall fields.

Inflation takes place along the trajectory where the vacuum expectation
value (VEV) of the inflaton obeys $S>M$ and where
$\overline G = G =0$. $V_0$ is completely flat in this direction. However,
since $V_0=\kappa^2 M^4$, SUSY is broken and the mass degeneracy
is lifted, such that the superfields $\widehat G$ and ${\widehat{\overline G}}$
encompass mass eigenstates of one fermion of mass $\kappa S$,
and two pairs of scalars
of mass squared $\kappa^2( S^2 \pm M^2)$ each. This induces
the Coleman-Weinberg radiative correction~\cite{CW,DSS}
\begin{eqnarray}
\label{VCW}
V_{\rm CW}=\frac{\kappa^4}{32\pi^2}\Big\{
(S^2 + M^2)^2 \ln \left(1+\frac{M^2}{S^2}\right)
+(S^2 - M^2)^2 \ln \left(1-\frac{M^2}{S^2}\right)
+2  M^4 \ln \frac{\kappa^2 S^2}{Q^2}
\Big\}
\,,
\end{eqnarray}
where $Q$ is a renormalisation scale. For notational convenience, here and in
the following it is always understood that we take the moduli of complex fields, for example $S\equiv |S|$, unless explicitly stated otherwise.
Added to $V_0$, the Coleman-Weinberg correction lifts the flatness and
forces $S$ to slowly roll towards zero. When the critical value
$S=M$ is reached, the waterfall fields acquire a negative mass square term
which forces them to assume the ${\rm U}(1)$-breaking VEV
$G=\overline{G}=M$, while $S$ is driven to zero. For realistic scenarios,
the ${\rm U}(1)$-group arises at an intermediate stage of the breaking of
the GUT-symmetry down to the Standard model group
$G_{\rm SM}={\rm SU}(3)_C \times {\rm SU}(2)_L \times {\rm U}(1)_Y$.
Possible candidates are
the baryon minus lepton symmetry $B-L$, the right-handed isospin or the
groups ${\rm U}(1)_X$ and ${\rm U}(1)_Z$ from the embedding
${\rm SO}(10)\supset {\rm SU}(5)\times {\rm U}(1)_X \supset {\rm SU}(3)_C \times {\rm SU}(2)_L \times {\rm U}(1)_X \times {\rm U}(1)_Z \supset G_{\rm SM} \times{\rm Z}_2$, see for example ref.~\cite{JeRoSa}. The spontaneous breakdown
of any combination of these ${\rm U}(1)$-symmetries at the waterfall
transition leads to the formation of local (gauged) cosmic strings.

Another interesting variant is $F_D$
inflation~\cite{GaPi,GaPaPi}, where the matter of the
Minimal Supersymmetric Standard Model (MSSM) is not charged under the broken
${\rm U}(1)$, giving rise to new long lived particles which may
loosen the gravitino bound discussed below. Moreover, in this scenario
the inflaton sector is tied to the MSSM-Higgs and to TeV-scale
right-handed neutrinos.
Note, however, that these additional couplings of the inflaton
alter the ``minimal'' form of the Coleman-Weinberg
potential~(\ref{VCW}). Nonetheless, the results presented here, in particular
that the string network allows for a blue spectral index, are also
qualitatively applicable to the $F_D$-model, and a quantitative study
constraining the model parameters, which might also be accessible at collider
experiments, can easily be performed using the methods applied here.

Assuming that supersymmetry is local,
supergravity gives rise to corrections of the following form
\begin{equation}
\label{V:SUGRA}
V_{\rm SUGRA} = c_H^2 H^2 S^2 + 
32 \pi^2\kappa ^2 M^4 \frac{S^4}{m_{\rm pl}^4}
+\dots\,,
\end{equation}
where $H^2= 8\pi \kappa^2 M^4 /(3 m_{\rm pl}^2)$ is the squared Hubble rate
during inflation and $m_{\rm pl}=1.22 \times 10^{19} {\rm GeV}$ denotes
the Planck mass.
While the term $\propto S^4$ is uniquely determined when only allowing
for renormalisable terms in the K\"ahler potential, the
$c_H^2$-term can be present for a ``non-minimal'' K\"ahler potential.
If $c_H^2$ is not imposed to be zero, for example by some
symmetry~\cite{cH2zero}, it is expected to
be of order one, which may incline the potential to an extent that
it becomes unsuitable for slow-roll inflation; an observation which is
known as the $\eta$-problem~\cite{CLLSW,DRT}.
For this reason, the minimal SUGRA-case with $c_H^2=0$ is most often considered
in the literature.
For the studies in this paper, we use the potential~(\ref{V:SUGRA}),
but note that also the
term $\propto S^4$ gets a correction factor $(1+O(c_H^2))$, which in turn may
again be modified by additional nonrenormalisable corrections which are
theoretically undetermined.
It has been pointed out recently that a
careful choice of the parameter $c_H^2$ brings the scalar spectral index $n_s$
into accordance with its central value determined from the WMAP3
data~\cite{BasShaKi} for models without cosmic strings. We shall return to this issue in section~\ref{sec:nonmin}.

The curvature-induced correction for SUSY in de Sitter background
can be derived to be~\cite{Garbrecht:2006df}
\begin{equation}
\label{V:R}
V_{R}=-\frac{3}{8\pi^2} H^2 \kappa^2 S^2 \ln \frac{\kappa^2 S^2}{Q^2}
\,.
\end{equation}
Since supersymmetry is not protected by non-renormalisation theorems in curved
space, the derivative of this contribution with respect to $S$ depends on
the cutoff scale $Q^2$. However, it turns out that for reasonable choices of
$Q$ between $\sqrt{\kappa} M$ and $m_{\rm Pl}$, the curvature correction may only be significant for large values of $\kappa$, where the contribution of strings to the power spectrum is too large to accord with observation.

Additional corrections from soft SUSY breaking turn out to be negligible,
except for the tadpole term
\begin{equation}
V_{\rm TP}=2 \kappa M^2 a_S{\rm Re}[S]\,,
\end{equation}
where $a_S$ should be of TeV-scale. The precise value of this parameter is
theoretically undetermined, and moreover, its effect also depends on the
phase of $a_S S$. It turns out that it may become important only for low
values, $\kappa \stackrel{<}{{}_\sim} 5 \times 10^{-4}$~\cite{sesh2,GaPaPi}.

Putting everything together, the full $F$-term inflationary potential is
given by
\begin{equation}
V=V_0+V_{\rm CW}+V_{\rm SUGRA}+V_R +V_{\rm TP}\,.
\end{equation}
In the subsequent numerical analysis, we study various scenarios arising as
special cases of this generic potential.

Some analytic understanding of the dependences between the
parameters can be gained from the following approximation.
We introduce the canonically
normalised inflaton $\sigma$ by making the choice of phase
$\sigma = \sqrt 2 {\rm Re}[S]$ and ${\rm Im}[S]=0$.
A simple estimate for the spectral index $n_{s}$ can be obtained for large
values $\kappa (\sim 10^{-2})$ by neglecting all
corrections other than
\begin{equation}
V=V_0+V_{\rm CW}=V_0+\frac{1}{8\pi^2} \kappa^4 M^4 \ln \frac{\kappa \sigma}{\sqrt 2 Q}\,.
\end{equation}
The number of ${\rm e}$-foldings between the time $t_{\rm e}$ at the
horizon exit of the scale  $k=0.05\,{\rm Mpc}^{-1}$
and the end of inflation
$t_{\rm c}$, where the inflaton reaches the critical value
$\sigma_{\rm c}=\sqrt 2 M$,
triggering the waterfall-phase transition, is given by
\begin{equation}
N_{\rm e}=\int_{t_{\rm e}}^{t_{\rm c}} dt\,H
=\frac{8\pi}{m_{\rm pl}^2} \kappa^2 M^4 
\int_{\sigma_{\rm c}}^{\sigma_{\rm e}} d\sigma \,
\left(\frac{\partial V}{\partial \sigma}\right)^{-1}
\approx \frac{32 \pi^3}{\kappa^2 m_{\rm pl}^2} \sigma_{\rm e}^2
\,,
\end{equation}
where we have used the slow-roll approximation
$3H \partial \sigma/\partial t=-\partial V/\partial \sigma$ and have neglected
the lower boundary term of the integral. We then find the
slow-roll parameter
\begin{equation}
\eta\big|_{N_{\rm e}}= \frac{m_{\rm pl}^2}{8\pi}
\frac{V^{\prime\prime}}{V}\Big|_{N_{\rm e}}=-\frac{1}{2 N_{\rm e}}\,,
\end{equation}
and consequently, when omitting the contribution from
the slow roll parameter $\epsilon$, which is negligible in these models,
\begin{equation}
\label{n_s:est}
n_s=1-2\eta=1-\frac{1}{N_{\rm e}} \approx 0.98
\,,
\end{equation}
for $N_{\rm e}\approx 50$.
Furthermore, the scale of inflation can now be determined by imposing the
observed amplitude of the primordial power spectrum,
\begin{equation} \label{power}
\sqrt{P_{\cal R}(k)}
=\frac{2^\frac 72 \sqrt \pi}{\sqrt 3 m_{\rm pl}^3}
\frac{V^\frac 32 (\sigma)}{\partial V /\partial \sigma}
\Bigg|_{\sigma= \sigma_{\rm e}}
=\frac{32 \pi}{\sqrt 3}\sqrt N_{\rm e} \frac{M^2}{m_{\rm pl}^2}\approx  4.54 \times 10^{-5}
\,,
\end{equation}
from which it follows that the symmetry breaking scale is close
to the Grand Unified scale, $M\approx4\times 10^{15} {\rm GeV}$. For
$P_{\cal R}(k)$, we have taken here the value determined by the standard
six parameter fit (the basic set of four plus $\{ \log(10^{10}P_{\cal R}),n_{s} \}$), see section~\ref{MCMC}.

\begin{figure}[t]
\begin{center}
\epsfig{file=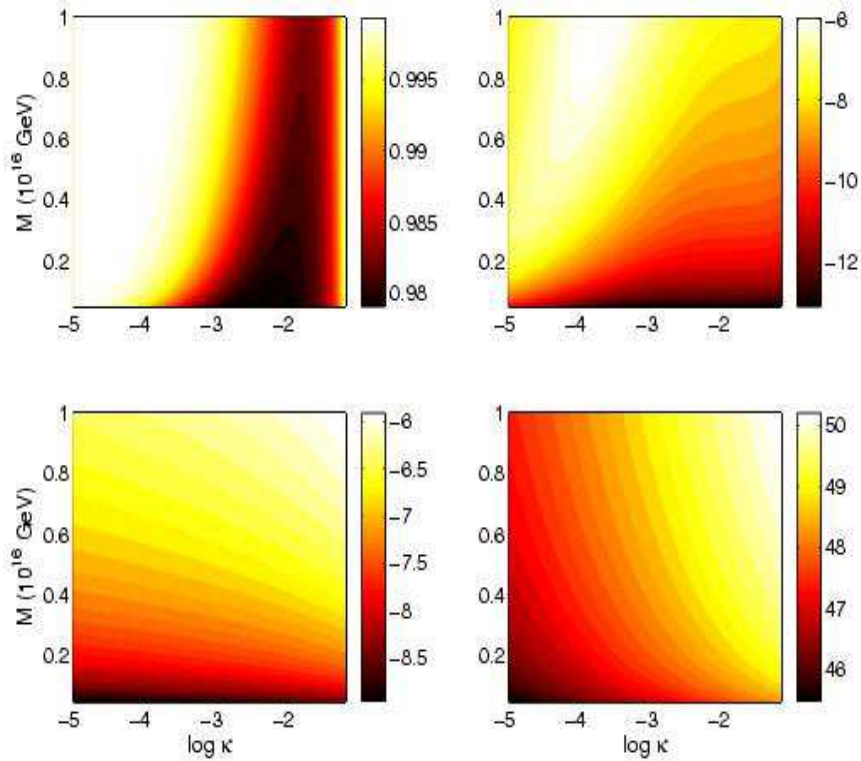,height=4.0in,width=4.5in}
\end{center}
\caption{The computed values of $n_{s}$ (top-left), $\log P_{\cal R}$ (top-right) $\log G\mu$ (bottom-left) and $N_e$ (bottom-right) as a function of $F$-term model parameters $\kappa$ and $M$ for $T_{\rm R}=10^{9}{\rm GeV}$ and g=0.7.}
\label{figure:Mkappa}
\end{figure}

We are considering the case where the gauge group is ${\rm U}(1)$ and therefore cosmic strings are expected to form during the waterfall transition. If the string network evolves towards a self-similar scaling regime as observed in simulations, then the initial conditions are unimportant and the only important parameter is the dimensionless string tension $G\mu$. This can be computed from the model parameters $\kappa$ and $M$ assuming that the strings do not have superconducting currents and results for the Abelian-Higgs model~\cite{HHT} can be applied. In ref.~\cite{JP1}, it was suggested that one could use 
\begin{equation}
\label{nonBogol}
G\mu=2\pi\left({M\over m_{\rm pl}}\right)^2\epsilon(\beta)\,,
\end{equation}
where $\beta=\kappa^2/(2g^2)$, $g$ is the gauge coupling assumed to be 0.7 based on Grand Unification and $\epsilon(\beta)=1.04\beta^{0.195}$ for $\beta>10^{-2}$ while $\epsilon(\beta)=2.4/\log(2/\beta)$ for $\beta<10^{-2}$.

In our subsequent analysis, we go beyond these simple estimates, which requires us to compute $n_{s}$ and $G\mu$ as functions of arbitrary $\kappa$ and $M$. We define $P_{\cal R}$ and $n_{s}$ at $k=0.05{\rm Mpc}^{-1}$.

We parameterize the number of e-foldings at the time when the scale $k=0.05{\rm Mpc}^{-1}$
crosses the horizon, $N_e$, using the unknown reheat temperature $T_{R}$ which is given by~\cite{BasShaKi}
\begin{equation}
N_{\rm e}\approx
50+\frac 13 \log \frac{T_{\rm R}}{10^9 {\rm GeV}}+
\frac 23 \log \frac{\sqrt \kappa M}{10^{15} {\rm GeV}}
\,.
\end{equation}
This relation is obtained by assuming that inflation is followed by
a matter dominated epoch of coherent oscillations, then a
radiation dominated epoch with the initial reheat temperature $T_{\rm R}$,
which lasts until matter-radiation equality, and eventually through matter domination until the present epoch.

The available constraints on $T_{\rm R}$ arise from the requirements that, within local
supersymmetry, gravitinos must not be overproduced~\cite{Gravitino} and that
successful baryogenesis takes place. Avoiding the gravitino problem gives
$10^{10}\,{\rm GeV}$ as a
conservative upper bound for $T_{\rm R}$, but depending on the gravitino mass
and its branching ratio into hadronic decays
constraints on the temperature as low as $10^{6}\,{\rm GeV}$ may
be in order. The bound on the reheat temperature may be relaxed, when
heavy particles decay at late times and release entropy, a feature which
is naturally invoked within the $F_D$-model~\cite{GaPi,GaPaPi}.
Scenarios for leptogenesis require lower bounds
of $T_{\rm R} \stackrel{>}{{}_\sim} 2\times10^9\,{\rm GeV}$ for non resonant thermal
leptogenesis~\cite{BuPeYa}, $0.3\,{\rm TeV}$ for resonant thermal
leptogenesis~\cite{ResLep} or
$3\times10^7\,{\rm GeV}$ when right handed neutrinos
are generated directly through the decay of the inflaton~\cite{sesh2}.
Note that in the range $0.3\,{\rm TeV} \leq T_{\rm R} \leq 10^{10}\,{\rm GeV}$,
the number of e-foldings varies just by two.
In section~\ref{sec:fterm_res}, we also study the effect a variation of $T_{\rm R}$
within the above bounds has on our results and find it to be small.
Throughout the rest of our analysis, we therefore choose the
value $T_{\rm R}=10^9\,{\rm GeV}$,
since lower reheat temperatures
require unnaturally small couplings of the inflaton-waterfall sector
to the Minimal Supersymmetric Standard Model (MSSM) matter.

\begin{figure}[t]
\begin{center}
\epsfig{file=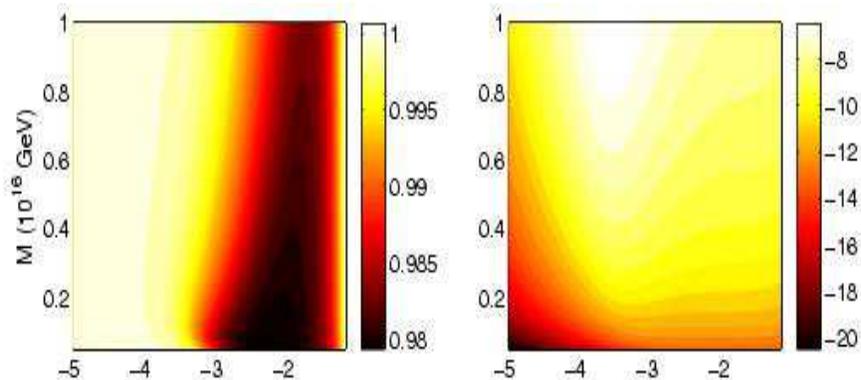,height=2.0in,width=4.5in}
\end{center}
\caption{The computed values of $n_{s}$ (left) and $\log P_{\cal R}$ (right) including the curvature $V_{R}$, and tadpole corrections, $V_{TP}$, to the $F$-term model for $a_S=1{\rm TeV}$, $T_{\rm R}=10^{9}{\rm GeV}$ and $g=0.7$.}
\label{figure:TADCURV}
\end{figure}

Including the full potential and performing a numerical study
for the minimal SUGRA-case $c_H^2=0$ and ignoring the contributions to the potential $V_{R}$ and $V_{\rm TP}$, it turns out that $0.98$ is just a lower bound for the spectral index. For large $\kappa$, the spectrum is turned blue due to the SUGRA-corrections, whereas for small $\kappa$, the inflaton evolves so slow that a quasi scale-invariant spectrum is generated~\cite{BasShaKi,GaPaPi,sesh1,sesh2,JP1,JP2}. This is illustrated in the Fig.~\ref{figure:Mkappa} for $T_{\rm R}=10^{9}{\rm GeV}$ and $g=0.7$. We note that we have used the full formula for $n_{s}$ in terms of the slow-roll parameters $\eta$ and $\epsilon$, although the contribution from $\epsilon$ turns out to be negligible. Moreover, we have computed the running of the spectral index, $n_{\rm run}$, and the tensor-to-scalar ratio, $r$; we found that $|n_{\rm run}|<10^{-3}$ and $r<10^{-4}$ for all the values of $\kappa$ and $M$ in Fig.~\ref{figure:Mkappa}.
We already pointed out that we expect corrections at low $\kappa$, if we include $V_{\rm TP}$, which is illustrated in Fig.~\ref{figure:TADCURV} for $a_{S}=1{\rm TeV}$.
As discussed above, the value of $a_S$ is not determined theoretically.
Qualitatively, one expects that for lower $a_S$ the deviations from the $a_S=0$
case only become important for even lower values of $\kappa$. We have checked
for example, that the minima of $P_{\cal R}$ in Fig.~\ref{figure:TADCURV}
is moved about a factor of two further to the left when choosing $a_S=0.1\,{\rm TeV}.$
We have also
included the curvature correction $V_R$, which has no significant impact
since at large $\kappa$, where it might become important, as the potential
is dominated by the minimal SUGRA correction.


\subsection{$D$-term inflation}
\label{Dterm}

\begin{figure}[t]
\begin{center}
\epsfig{file=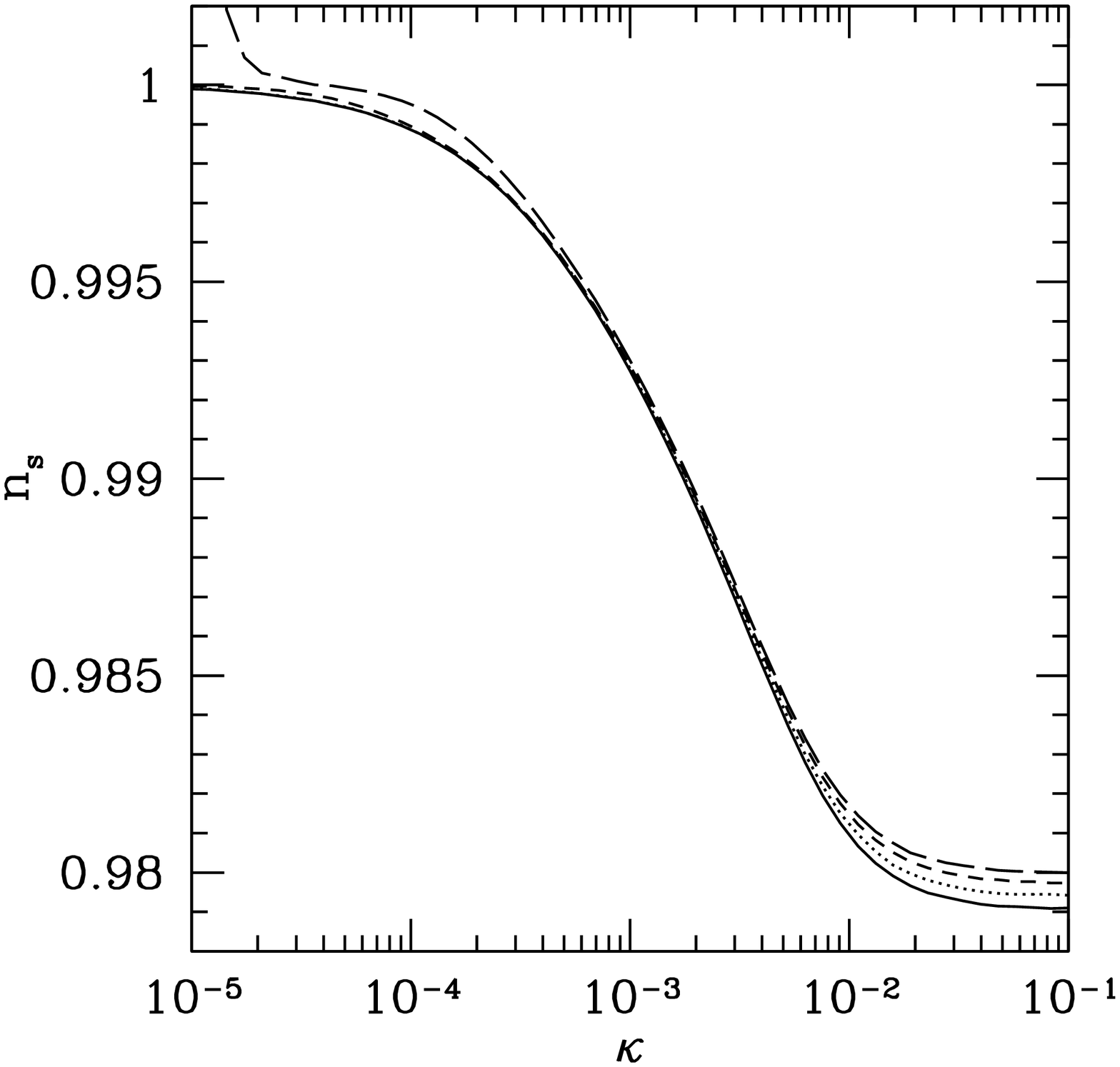,width=7cm}
\epsfig{file=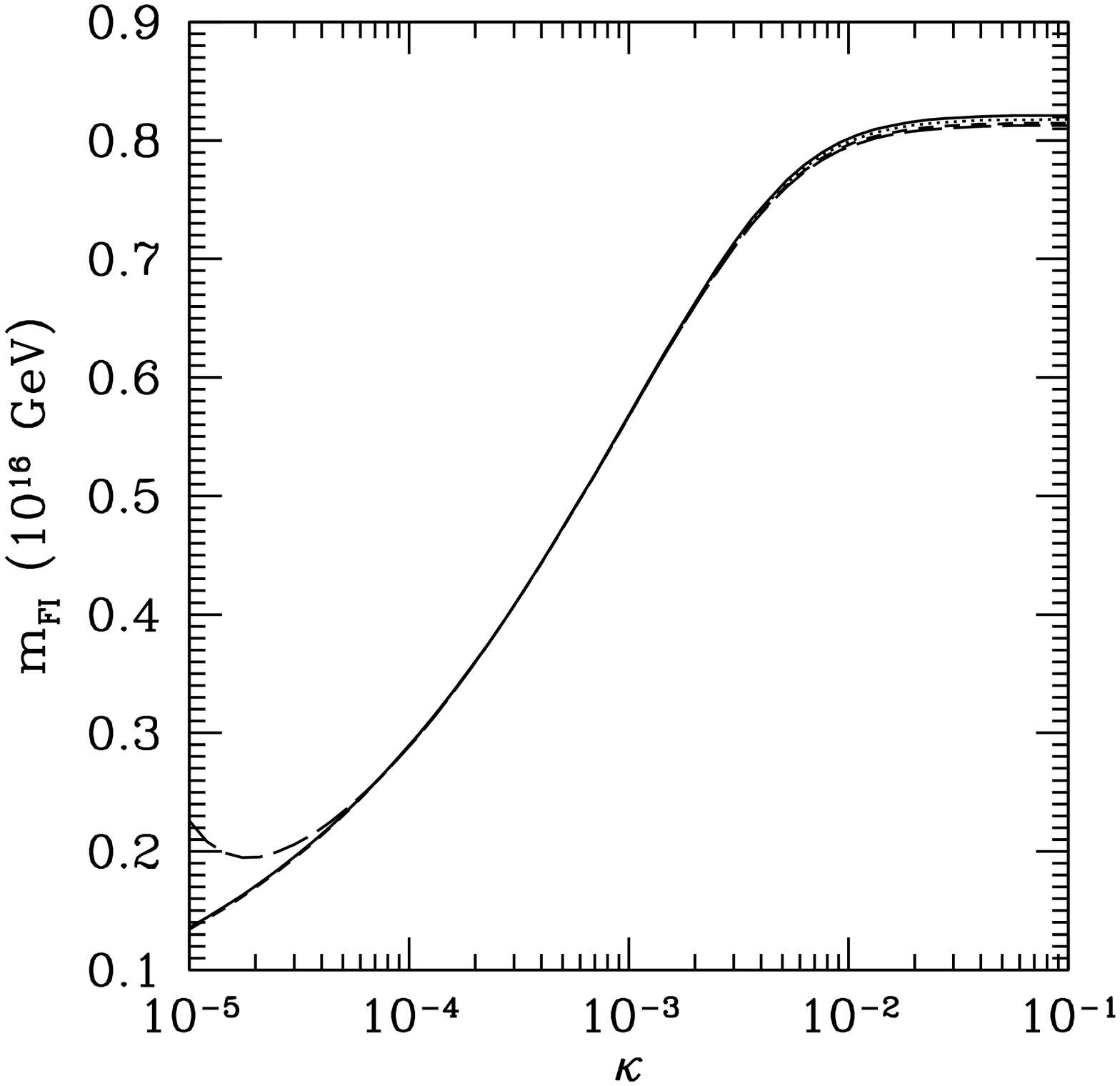,width=7cm}
\end{center}
\caption{The computed values of $n_{s}$ (left) and $m_{\rm FI}$ for a fixed $P_{\cal R}$, given by (\ref{power}). The solid line has the SUGRA parameter  $g=10^{-4}$, the dotted line $g=10^{-3}$ and the short-dashed line $g=10^{-2}$. Below $g=10^{-2}$ the observable properties are only very weakly dependent on $g$ for the range of $\kappa$ considered. The long-dashed line shows $g=0.05$, and for small values of $\kappa$ the induced values of $m_{\rm FI}$ and $g$ are increased.}
\label{figure:dterm2}
\end{figure}

The $D$-term model shares many features with $F$-term hybrid inflation.
It is implemented by the
superpotential~\cite{DTerminf}
\begin{equation}
W=\kappa \widehat S {\widehat{\overline G}} \widehat G\,,
\end{equation}
and by the $D$-term
\begin{equation}
D=\frac g2 \left(|G|^2 -|\overline G|^2 +m_{\rm FI}^2\right)\,,
\end{equation}
where $m_{\rm FI}$ denotes the Fayet-Iliopoulos mass.
These combine to give the tree-level scalar potential
\begin{equation}
V=\kappa^2\bigg[|\overline G G|^2 +|S \overline G |^2 +|S G|^2\bigg]
+\frac 12 D^2\,.
\end{equation}
Note that the $D$-term is also present in $F$-term models, but that it just imposes
$|\overline G|= |G|$ since $m_{\rm FI}=0$, while in the present case, it is
responsible for the spontaneous breakdown  of the ${\rm U}(1)$-gauge symmetry
after inflation which occurs at
$S^2=S_{\rm c}^2=\frac{g^2}{4\kappa^2} m_{\rm FI}^2$. Above this value for
$S$, $\overline G =G =0$, and  since $V_0=\frac{g^2}{8}m_{\rm FI}^4$
is non-zero, SUSY and therefore the mass-degeneracy between bosons and fermions
is broken.
The
mass eigenstates comprised in $\widehat G$ and ${\widehat{\overline G}}$
are one Dirac fermion of mass $\kappa s$ and two pairs of scalars with
mass squared $\kappa^2 s^2 \pm \frac{g^2}{4} m_{\rm FI}^2$, which
induce the the radiative correction
\begin{eqnarray}
V_{\rm CW}=\frac{1}{32\pi^2}\Big\{
(\kappa^2 s^2 + \frac{g^2}{4} m_{\rm FI}^2)^2 
\ln \left(1+\frac{g^2}{4 \kappa^2} \frac{m_{\rm FI}^2}{s^2}\right)
+(\kappa^2 s^2 - \frac{g^2}{4} m_{\rm FI}^2)^2 
\ln \left(1-\frac{g^2}{4 \kappa^2}\frac{m_{\rm FI}^2}{s^2}\right)
+\frac{g^4}{8} m_{\rm FI}^4 \ln \frac{\kappa^2 s^2}{Q^2}
\Big\}
\,,
\end{eqnarray}
where within minimal SUGRA
\begin{equation}
s=S{\rm e}^{8\pi\frac{S^2}{m_{\rm Pl}^2}}\,.
\end{equation}
Note that a corresponding correction also applies to the Coleman-Weinberg
potential in the $F$-term model, but it is negligible when compared with the
minimal SUGRA correction in that case. In turn, the minimal SUGRA correction and the possible $c_H^2$ term do not occur in the $D$-term model due to the 
vanishing of the $F$-terms.
When ignoring SUGRA or assuming its absence, we can simply set $s=S$.
One can then prove that
the VEV of the inflaton at horizon exit $\sigma_{\rm e}$ is proportional to
$g$. For a given $P_{\cal R}$, the induced values of $m_{\rm FI}$ and
$n_{s}$ are then independent of $g$. Numerically, it turns out that
the approximation $s=S$ and therefore the degeneracy in the parameter $g$
is good for $g\stackrel{<}{{}_\sim}0.1$ and where the
parameter $\kappa$ is within the allowed range, see Fig.~\ref{figure:dterm2}.

We note that a tadpole correction does not occur in the $D$-term model, but
there is the curvature term~(\ref{V:R}). Since this only contributes
significantly for large values of $\kappa$ where there is 
a large contribution of cosmic strings to the CMB-temperature fluctuations, we do not consider it any further here.

The dimensionless string tension in these models is given by $G\mu=2\pi(m_{\rm FI}/m_{\rm pl})^2$, since the strings satisfy the Bogomol'nyi bound. Moreover, the relationship between $N_{\rm e}$ and $T_{\rm R}$ is modified to 
\begin{equation}
N_{\rm e}\approx
50+\frac 13 \log \frac{T_{\rm R}}{10^9 {\rm GeV}}+
\frac 23 \log \frac{\sqrt g m_{\rm FI}}{10^{15} {\rm GeV}}
\,.
\end{equation}

\begin{figure}[t]
\begin{center}
\epsfig{file=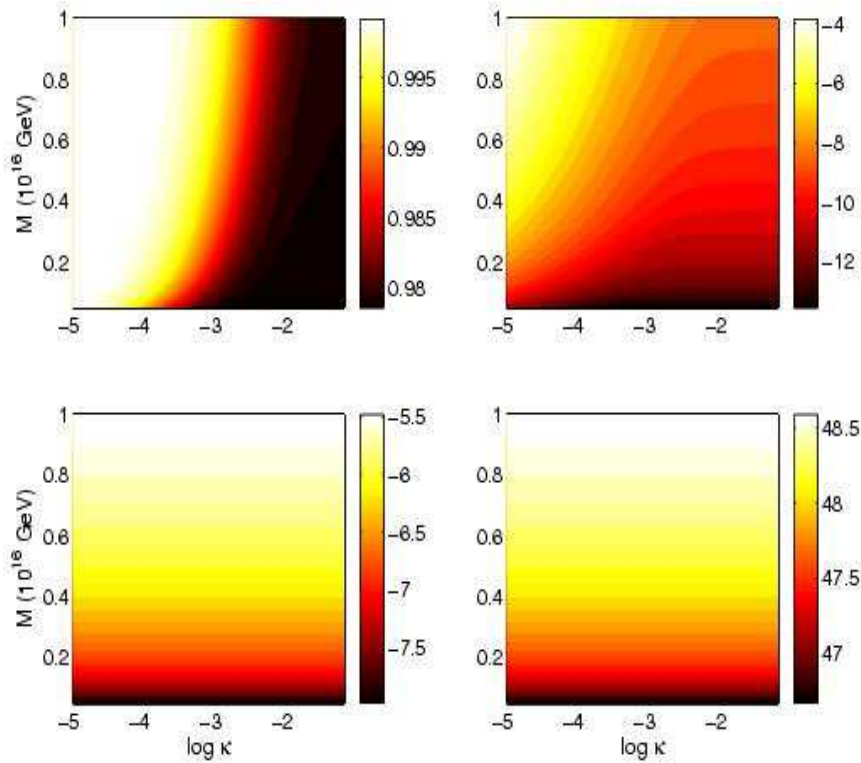,height=4.0in,width=4.5in}
\end{center}
\caption{The computed values of $n_{s}$ (top-left), $\log P_{\cal R}$ (top-right), $\log G\mu$ (bottom-left) and $N_e$ (bottom-right) as a function of $D$-term model parameters $\kappa$ and $M_{\rm FI}$ for $g=10^{-3}$, $T_{\rm R}=10^{9}{\rm GeV}$ and $s=S$.}
\label{figure:dterm}
\end{figure}

As in the $F$-term case we have computed the important observational quantities ignoring the $V_R$ term, and the results are presented in Fig.~\ref{figure:dterm} using $g=10^{-3}$, $T_{\rm R}=10^{9}{\rm GeV}$ and $s=S$. In accordance with the above discussion, these results are independent of the parameter $g$ if we ignore SUGRA, and are also a good approximation for the minimal SUGRA case when $g$ is small ($\stackrel{<}{{}_\sim} 0.1$).

\subsection{Cosmic string power spectrum}
\label{Sspec}

Work in the late 1990s led to strings being excluded as the primary source of cosmic fluctuations~\cite{PST,ABR,Knox,CHMa}, since they cannot reproduce the observed peak structure of the power spectrum due to decoherence~\cite{ACFMa}, and the maximum of the spectrum is located at the multipole moment $\ell\approx 500$, since the fluctuations are created at the scale corresponding to the correlation length~\cite{ACFMb}. More recent work has improved the accuracy of the string spectrum~\cite{PV,bevis,Landriau}, but since the string contribution to the power spectrum is at most 10\%, substantially less accurate predictions are necessary than for the dominant adiabatic component.

\begin{figure}[htbp]
\epsfig{file=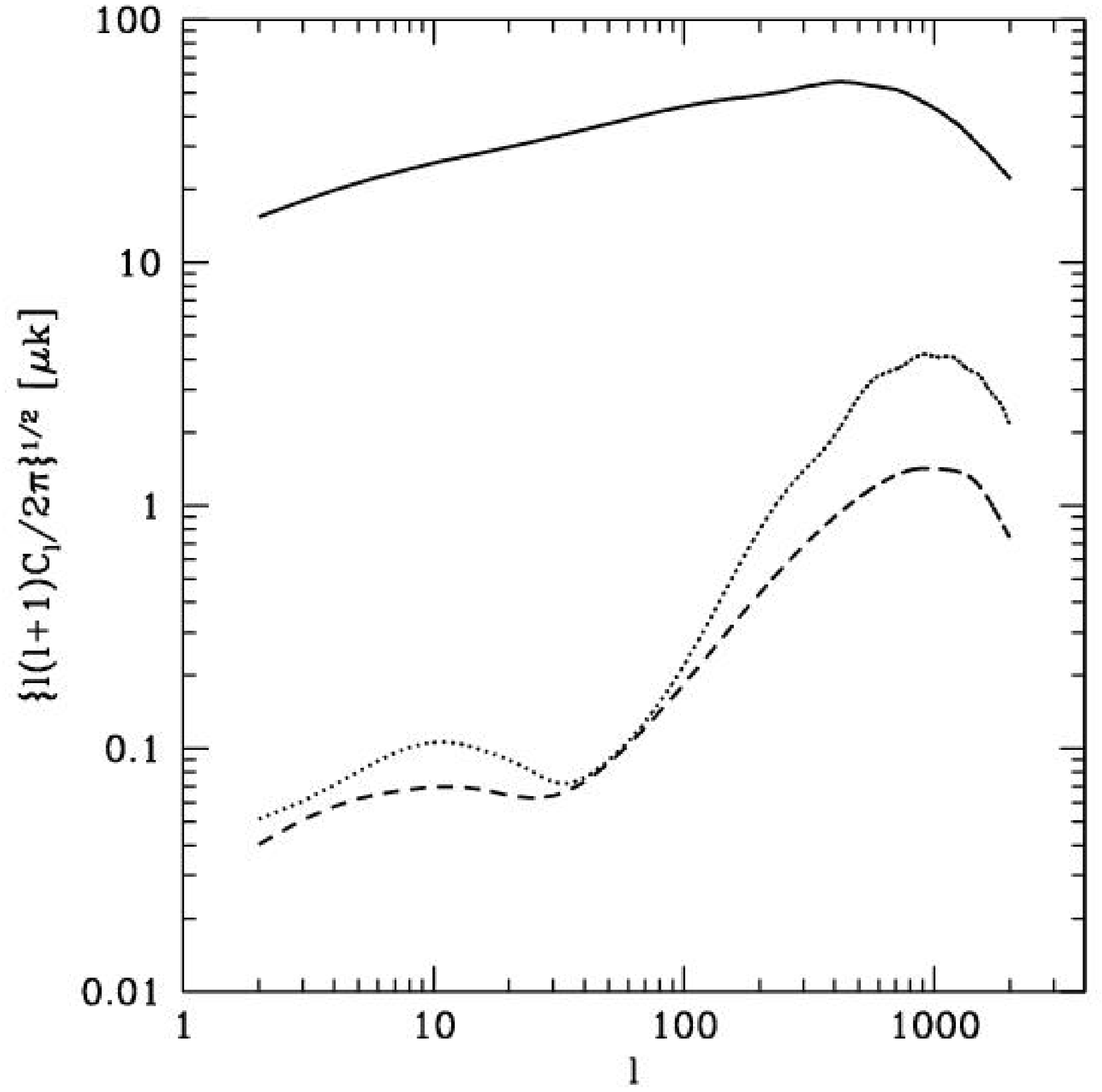,width=8cm}
\epsfig{file=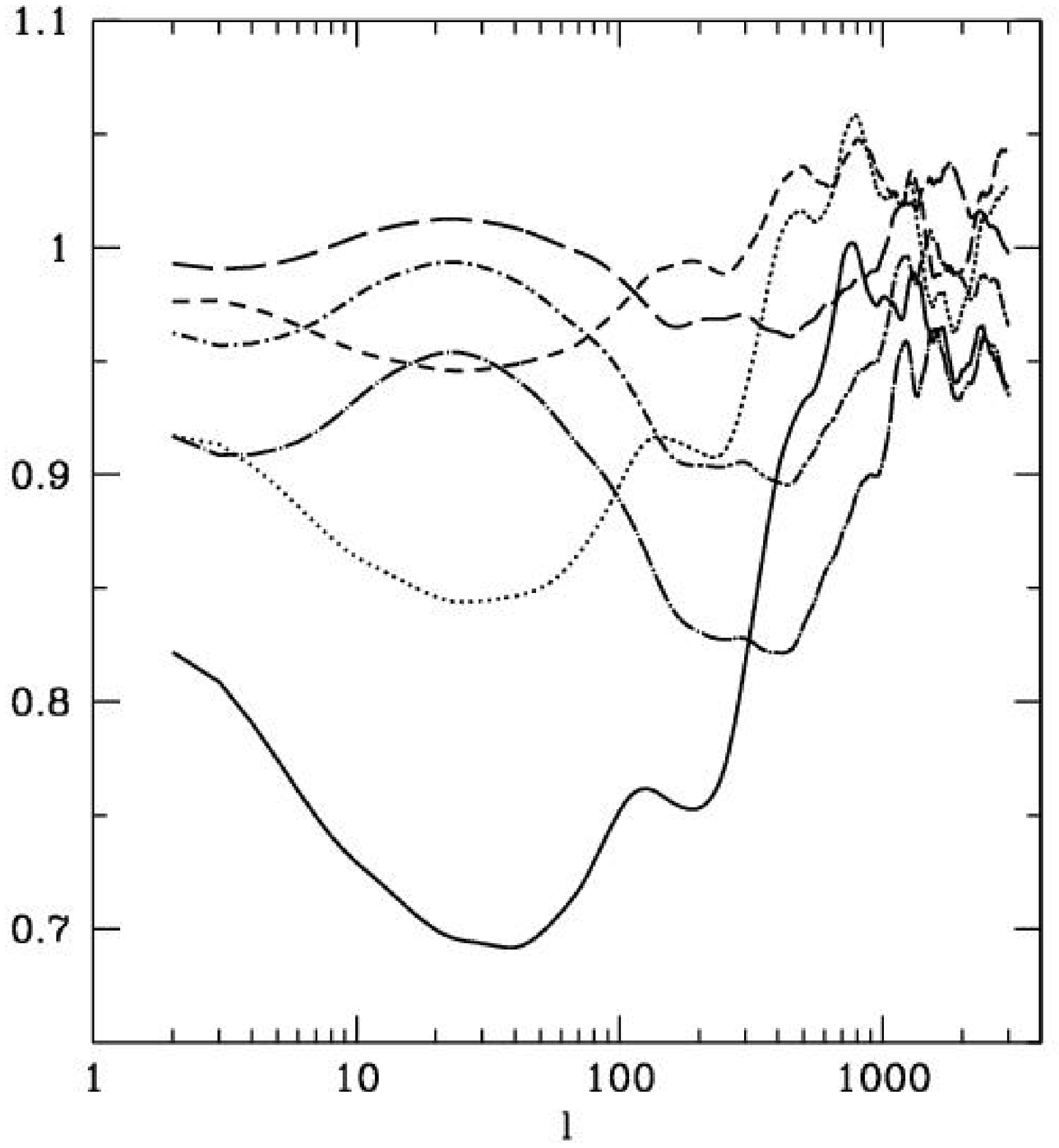,width=8cm}
\caption{Spectrum of anisotropies predicted by the cosmic string model used in this analysis normalized to COBE. The solid line is the temperature anisotropy, the dotted line is the $E$-mode polarization and the dashed line if the $B$-mode polarization. In  the right panel we show the ratio of the temperature spectra for $\beta_{r}=1$ (solid), $\beta_{r}=1.3$ (dot), $\beta_{r}=1.6$ (short dash), $\beta_{r}=2.2$ (long dash), $\beta_{r}=2.5$ (dot-short dash), $\beta_{r}=2.8$ (dot-long dash) compared to $\beta_{r}=1.9$.}
\label{fig:stringspec}
\end{figure}

We have used the model described in ref.~\cite{PV} which is an adaptation of that first proposed in ref.~\cite{ABR}. It models the string network in the radiation era as a set of line segments with a given length $\xi_{r}=0.26$ relative to the horizon and rms velocity $\langle v_{r}^2\rangle^\frac 12=0.65$, where the functional extrapolation from the radiation era to the matter era can be found in ref.~\cite{MS}. The adaptation of ref.~\cite{PV} also includes the effects of string ``wiggles''\cite{carter,vilenkin} via the parameter $\beta=\sqrt{\tilde\mu/\tilde T}$ where $\tilde\mu$ and $\tilde T$ are the mass per unit length and tension of the wiggly strings. The value of $\beta$ in the radiation and matter eras has been estimated to be $\beta_{r} \approx 1.9$  and $\beta_{m} \approx 1.5$.  The interpolation between the two epochs is achieved using the function $\beta(\eta)=1+(\beta_{r}-1)a/(\eta \dot{a})$, where an overdot represents the derivative with respect to the conformal time $\eta$. In Fig.~\ref{fig:stringspec} we plot the angular power spectra for the temperature and polarization predicted by cosmic strings with $\beta_{r}=1.9$. This spectrum was computed by averaging over 400 string network realisations and using the WMAP best fit cosmological parameters $\Omega_{\rm b}h^2=0.0223$, $\Omega_{\rm m}h^2=0.127$, $h=0.73$ and $\tau_{R}=0.088$. $\Omega_{\rm b}$, $\Omega_{\rm c}$ and $\Omega_{\rm m}=\Omega_{\rm b}+\Omega_{\rm c}$ are the densities of baryons, cold dark matter and matter defined relative to critical, and $h=H_0/(100{\rm km}\,{\rm sec}^{-1}\,{\rm Mpc}^{-1})$. We also show the effect of changing the wiggliness parameter on the temperature power spectrum by plotting the ratio of the spectrum compared to $\beta_{r}=1.9$ for a variety of values of $\beta_r$. We see that if $1.3\le\beta_{r}\le 2.8$ the effect of changing $\beta_r$ is at most a 20\% effect. We find similar size modifications for sensible variations in the other two parameters, $\xi_{r}$ and $\langle v^2\rangle_{r}^{1/2}$.

\subsection{MCMC analysis}
\label{MCMC}

Since computation of the string power spectrum takes $\approx 24$ hours for each set of cosmological parameters it is not feasible to do this at every step of a Markov-Chain-Monte-Carlo (MCMC) analysis. However, since the string power spectrum is likely to be only around 5\% of the total, if the string spectrum varies by less than, say, 10-20\% in the range of parameters allowed by the WMAP data assuming only adiabatic perturbations, the error introduced by assuming that the string spectrum is unchanged by any variation in the cosmological parameters is less than the 1\% accuracy claimed by codes such as {\tt CMBFAST}~\cite{Seljak} and {\tt CAMB}~\cite{LC}. This would allow very fast MCMC analysis using a single extra parameter $G\mu$ which normalizes the string power spectrum.

\begin{figure}[t]
\epsfig{file=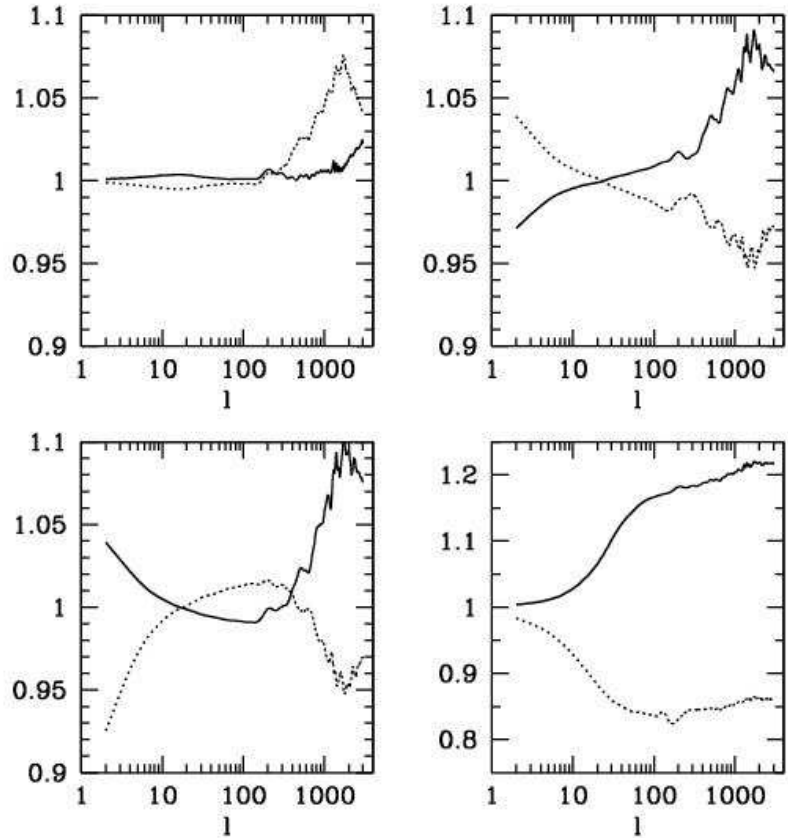,width=12cm}
\caption{The effect of different cosmological parameters on the temperature power spectrum predicted by the fiducial cosmic string model. In each panel we plot the ratio of the spectrum predicted by the the WMAP best fit parameters (with $\beta_{r}=1.9$) compared to upper and lower $3-\sigma$ bounds of these parameters . In the top-left panel, for example, we plot the $3-\sigma$ upper bound on $\Omega_b h^2=0.0244$ (solid) and the $3-\sigma$ lower bound on $\Omega_b h^2=0.0196$ (dotted). In the top right panel we plot $\Omega_m h^2=(0.148, 0.100)$, the bottom-left panel shows $h=(0.82, 0.61)$ and the bottom-right panel $\tau_R=(0.172, 0)$.}
\label{fig:ttcomp}
\end{figure}

In Fig.~\ref{fig:ttcomp} we present the ratio of the string spectra for parameters which are $3-\sigma$ away, as defined by the constraints on adiabatic models from WMAP, from the fiducial spectrum. We see that there is, at most, a 20\% variation in the spectrum in each case. Therefore, assuming that the string model itself is correct, it appears safe to ignore the variation of cosmological parameters on the string spectrum. 

The MCMC analysis used the May 2006 version of {\tt COSMOMC}~\cite{LB} in order to create  chains which were used to estimate confidence limits on the cosmological parameters. The basic set of four parameters $\{ \Omega_{\rm b}h^2, \Omega_{\rm c}h^2, \tau_{\rm R}, \theta_{\rm A} \}$, where $\theta_{\rm A}$ is defined by the ratio of the sound horizon to the angular diameter distance at the redshift of recombination, were used in each case. In addition, we vary the set $\{ \log(10^{10}P_{\cal R}),n_{s}, G \mu \}$, or alternatively derive these parameters from an inflationary model of interest defined by $\{ M, \log \kappa, c_{H}^{2}, m_{\rm FI}, \log g, ..... \}$. In tables~\ref{FTermSummary} and~\ref{DTermSummary} of the appendix, there is, for comparison, a fit for the parameters $\{ \Omega_{\rm b}h^2, \Omega_{\rm c}h^2, \tau_{\rm R}, \theta_{\rm A},\log(10^{10}P_{\cal R}),n_{s} \}$; we refer to this as the standard six parameter fit, there and elsewhere in the text.

For the analysis of section~\ref{sec:blue} we use three parameters $\{\log(10^{10}P_{\cal R}),n_s,G\mu\}$ to describe the power spectrum. For the analyses of the $F$-term and $D$-term models, we first tried using the model parameters $\{M,\log \kappa\}$ or $\{m_{\rm FI},\log \kappa\}$ and computed $\{\log(10^{10}P_{\cal R}),n_s,G\mu\}$ from them. However, since the range of these parameters allowed by the data is a very narrow degenerate line, the acceptance rate for the Markov Chains was very low. As an alternative we found that using $\log(10^{10}P_{\cal R})$ and $\log \kappa$ as the parameters allowed for rapid convergence of the chains. The values of $M$ (or $m_{\rm FI}$), $G\mu$ were then computed as derived parameters. Note that this approach also corresponds to a six parameter fit. For the simplest models, one needs one further piece of information, $N_{\rm e}$ and this was parameterized by the reheat temperature, $T_{\rm R}$. In addition, where necessary $c_H^2$ and $\log g$ were used as additional parameters.

We use data from the 3rd year observations from WMAP~\cite{Hinshaw,Page} (WMAP3 and WMAP1 are used to refer to the 3rd year and 1st year data) and three experiments which observe at substantially higher resolution than possible using WMAP. These are the Cosmic Background Imager (CBI)~\cite{Readhead}, the ArCminute Bolometer ARray (ACBAR)~\cite{Kuo} and BOOMERANG~\cite{Piacentini,Jones,Montroy}.  The intrinsic flat priors, listed in Table \ref{flatpriors}, were chosen to be sufficiently broad to incorporate the lines of degeneracy known to exist within the space of parameters.

\begin{table}
\begin{center}
\begin{tabular}{|c|c|} \hline
Parameter & Prior \\ \hline
$\Omega_{\rm b} h^2 $ & (0.005, 0.1) \\ 
$\Omega_{\rm c} h^2 $ & (0.01, 0.99) \\ 
$\theta_{\rm A}  $ & (0.5, 10) \\ 
$\tau_{\rm R} $ & (0.01, 0.9) \\ 
$\log (10^{10} P_{\cal R})$ & (2.7, 5.0) \\ \hline
$n_{s} $ & (0.5, 1.5) \\ \hline
$\log \kappa$ & (-5.0, -0.3) \\
$\log(T_{\rm R} /10^{9} {\rm GeV})$ & (-6.0, 1.0) \\ \hline
$c_H^{2}$  & (-0.25, 0.03) \\ \hline
$\log g$ & (-2.0, 0.0) \\ \hline
\end{tabular}
\end{center}
\caption{\label{flatpriors} Table of flat priors. The notation $(a,b)$ for a particular parameter gives the lower and upper bounds allowed in the fit.}
\end{table}

\section{Results}
In this section, we present the results of the likelihood analyses for various
scenarios. The standard six parameter fit extended by the string tension
$G\mu$ is discussed in~\ref{sec:blue}, the $F$-term models with
minimal SUGRA in~\ref{sec:fterm_res}, $D$-term inflation in~\ref{sec:dterm_res}
and finally $F$-term inflation with non-minimal SUGRA in~\ref{sec:nonmin}.
All results of this section are summarized in tables~\ref{FTermSummary}
and~\ref{DTermSummary} in the appendix, so that they can easily be compared.

\subsection{Inclusion of strings allows blue power spectra}
\label{sec:blue}

\begin{figure}[htbp]
\epsfig{file=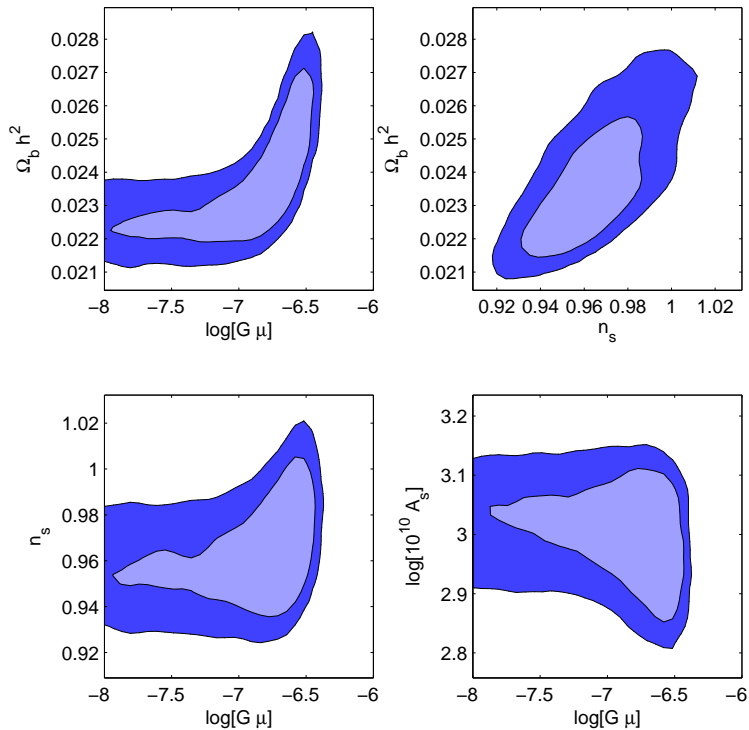,width=10cm}
\caption{Results for a 7 parameter fit to the CMB data. The contours are the 68\% and 95\% confidence levels in each of the 2D planes.}
\label{fig:7param}
\end{figure}

It was pointed out by the WMAP team~\cite{wmap} that the tight constraint on $n_{s}=0.951_{-0.019}^{+0.015}$ only applies to $r=0$, and that larger values are allowed for non-zero $r$. The SUSY hybrid inflation models under consideration here give rise to $r<10^{-4}$, but the string contribution to the power spectrum is similar in many ways to that for tensors and our investigation of this effect was first motivated by the possibility that there is a degeneracy between $n_{s}$ and $G\mu$ which would allow $n_{s}\approx 0.98$ to be accommodated more comfortably.

The results of performing a 7-parameter fit (the basic set of four plus $\{ \log(10^{10}P_{\cal R}),n_{s}, G \mu \}$) are presented in Fig.~\ref{fig:7param}. There is a $2-\sigma$ upper bound on $G\mu<3.0\times 10^{-7}$ (marginalized) which is compatible with previous estimates~\cite{pog,bevis,slosel}. It can be seen clearly that for $G\mu<10^{-7}$ lower values of $n_{s}=0.94-0.98$ are preferred, whereas for $G\mu>10^{-7}$ values as large as $n_{s}=1.02$ are within the $2-\sigma$ contour. Larger values of $n_{s}$ also require larger values of $\Omega_{\rm b}h^2$; if we fix $n_{s}=0.98$ and perform a 6 parameter fit then $G\mu=2.5 \times 10^{-7}$ is the best fitting value and $\Omega_{\rm b}h^2\approx 0.025$. This analysis appears to compatible with the intuition described above and suggests that a more detailed analysis of the more restricted parameter space for the $F$- and $D$-term models is warranted.

The computed value of $-2\log{\cal L}=11302.8$ for the likelihood of the 7 parameter fit compares to 11305.5 for a 6 parameter fit which corresponds to a $\Delta\chi^2\approx 2.7$. It is clear from this that any sensible Bayesian model selection criterion would not favour the 7 parameter model over that with 6 parameters. If we fix $n_{s}=0.98$ and fit for $G\mu$ as the sixth parameter, then we find that $-2\log{\cal L}=11303.0$ suggesting that this 6 parameter model gives an equally good fit.

If we impose $\Omega_{\rm b}h^2=0.020\pm 0.002$ as suggested by Big Bang Nucleosynthesis we find that $G\mu<2.2\times 10^{-7}$ (marginalized) and $n_{s}=0.953 \pm 0.015$.

\subsection{Minimal $F$-term models}
\label{sec:fterm_res}

\begin{figure}[htbp]
\epsfig{file=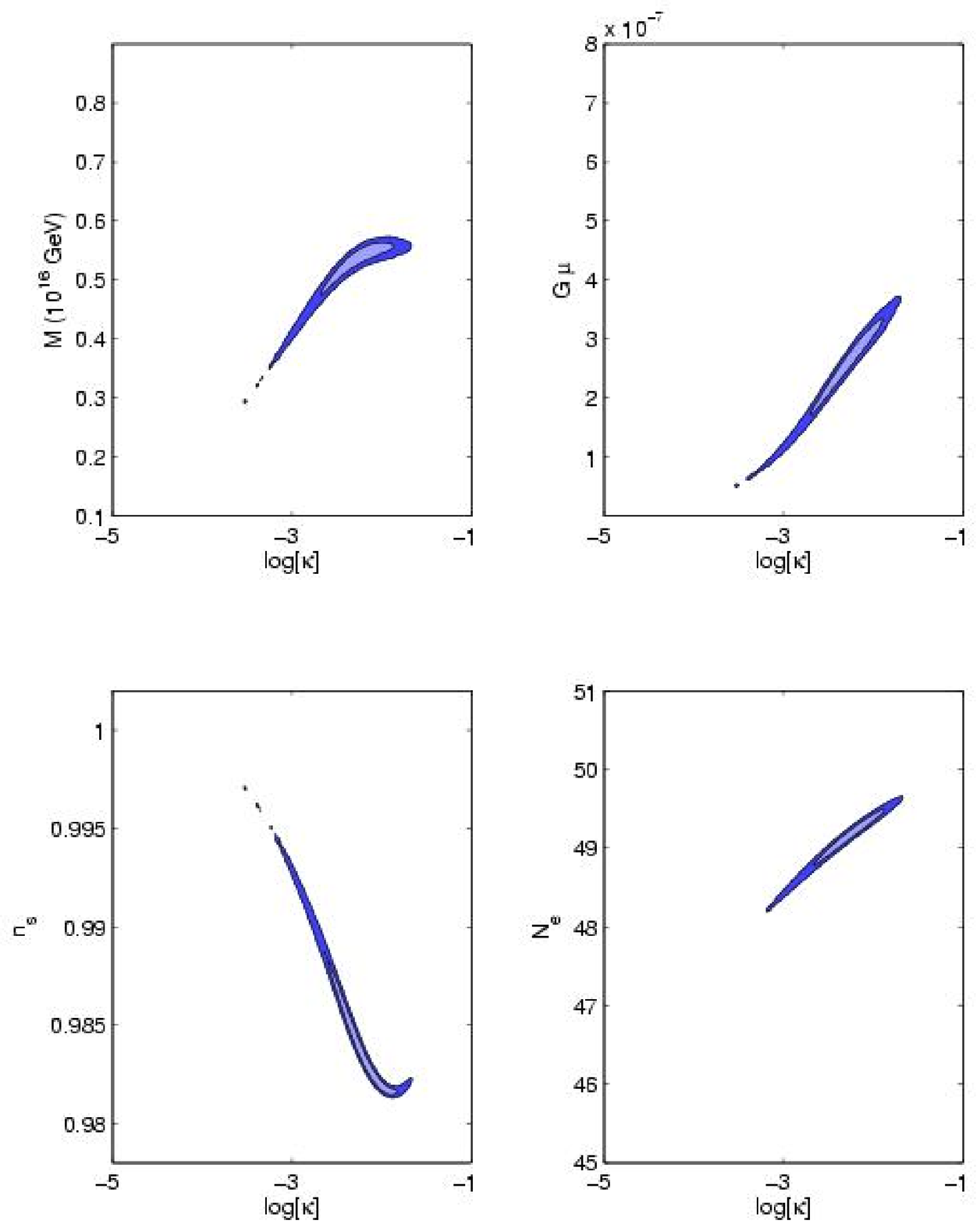,width=8.5cm}
\epsfig{file=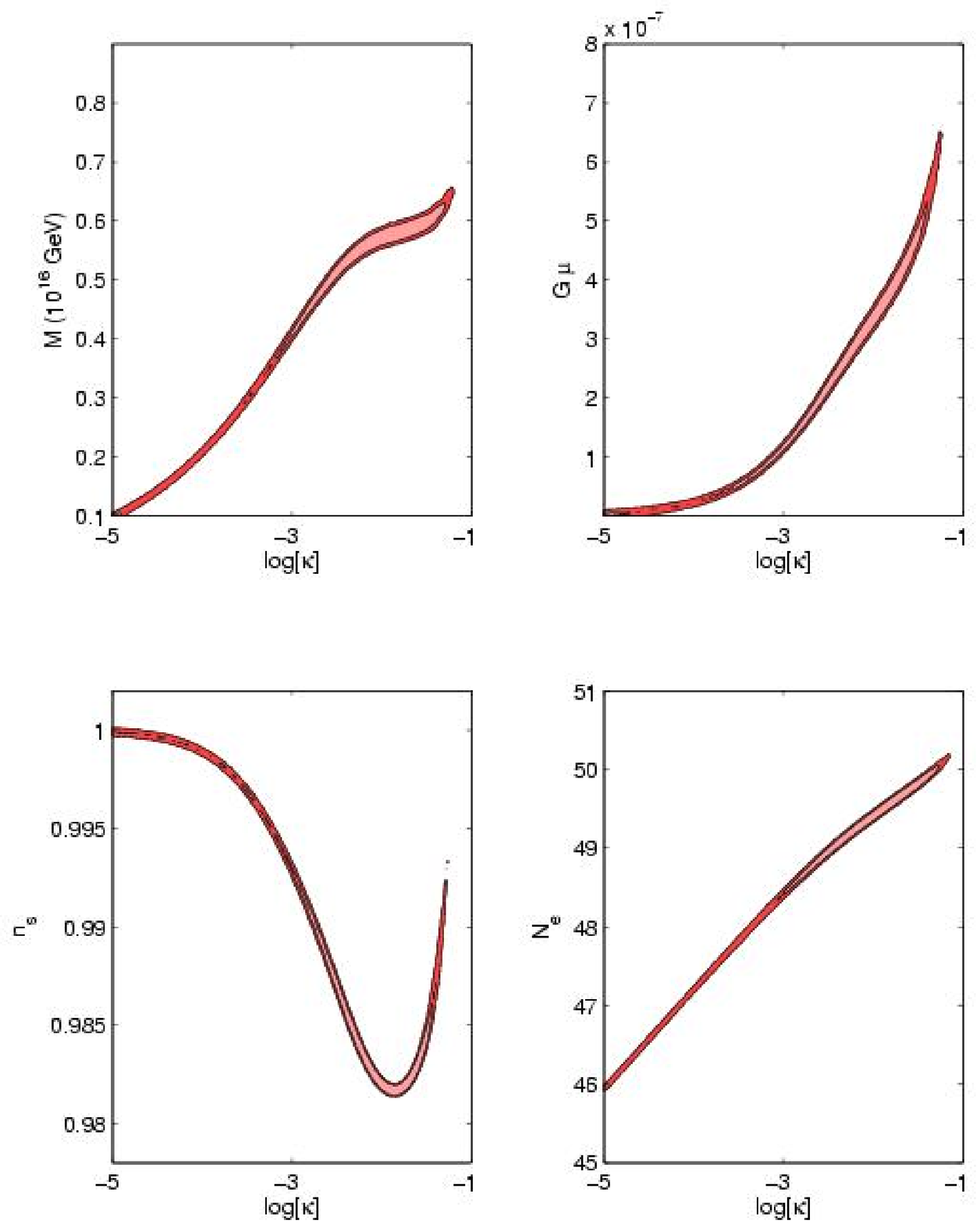,width=8.5cm}
\caption{Constraints on the $F$-term model with minimal SUGRA and $g=0.7$, $T_{\rm R} =10^9 {\rm GeV}$. On the left are the allowed regions when strings are included and on the right when they are not included. In the latter case the value of $G\mu$ that would be inferred by the preferred values of $\kappa$ and $M$ is presented. It is clear that the inclusion of the string component is critical to determining the constraints on $\kappa$ and $M$.
\label{figure:FTERMres}}
\end{figure}

The results of an analysis of minimal $F$-term models  in which we have varied $\kappa$ and $\log(10^{10}P_{\cal R})$ with $T_{\rm R}=10^9{\rm GeV}$ and $g=0.7$ including the string contribution (with $n_s$, $G\mu$, $M$ and $N_e$ as derived parameters) are presented in Fig.~\ref{figure:FTERMres}. We find that
$\log\kappa=-2.34\pm 0.38$
and $M=(0.518\pm 0.059)\times 10^{16}{\rm GeV}$, although there is a strong correlation between the two. The best fitting model has $-2\log{\cal L}=11303.3$, which is close to that of the standard 6 parameter fit.
The best fit value for the string tension is $G\mu=2.6\times 10^{-7}$, which corresponds to $6\%$ of the power spectrum amplitude at $\ell=10$. We find that $n_s= 0.985 \pm 0.004 $, $G \mu = (2.5 \pm 0.6) \times  10^{-7} $ and $\Omega_{\rm b}h^2 = 0.0255 \pm 0.0009$.

We have performed the same analysis but have left out the string contribution to the power spectrum. The results of this analysis are also presented in Fig.~\ref{figure:FTERMres}. The range of values in the $\kappa-M$ plane which are allowed in this (incorrect) analysis are much larger and for the best fit, $-2\log{\cal L}=11308.4$ which is significantly worse than that with strings included. If strings are included, large values of $\kappa$ are disallowed since they give rise to an excess string component, $G\mu>3\times 10^{-7}$.

\begin{figure}[htbp]
\epsfig{file=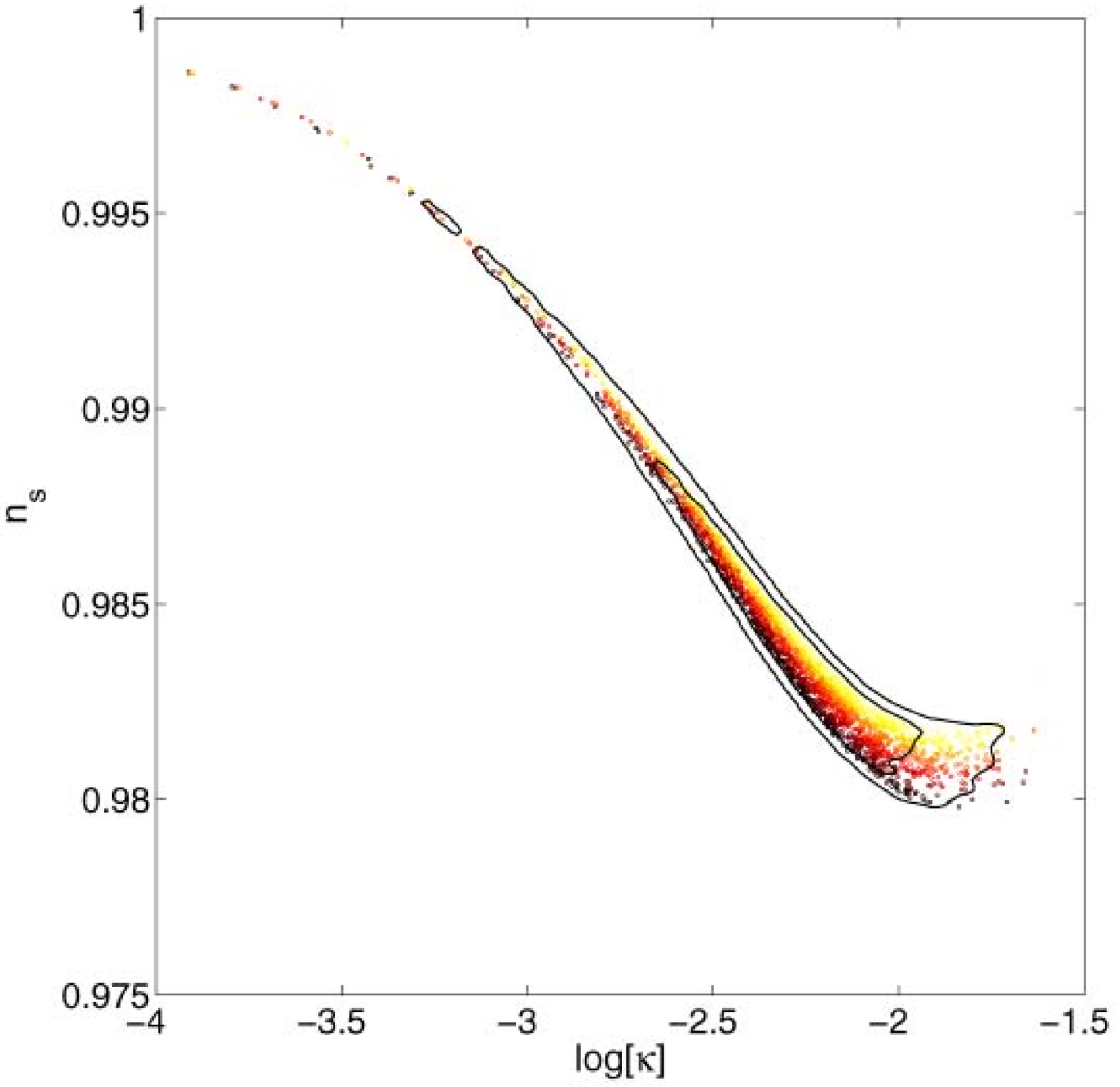,width=7.1cm}
\epsfig{file=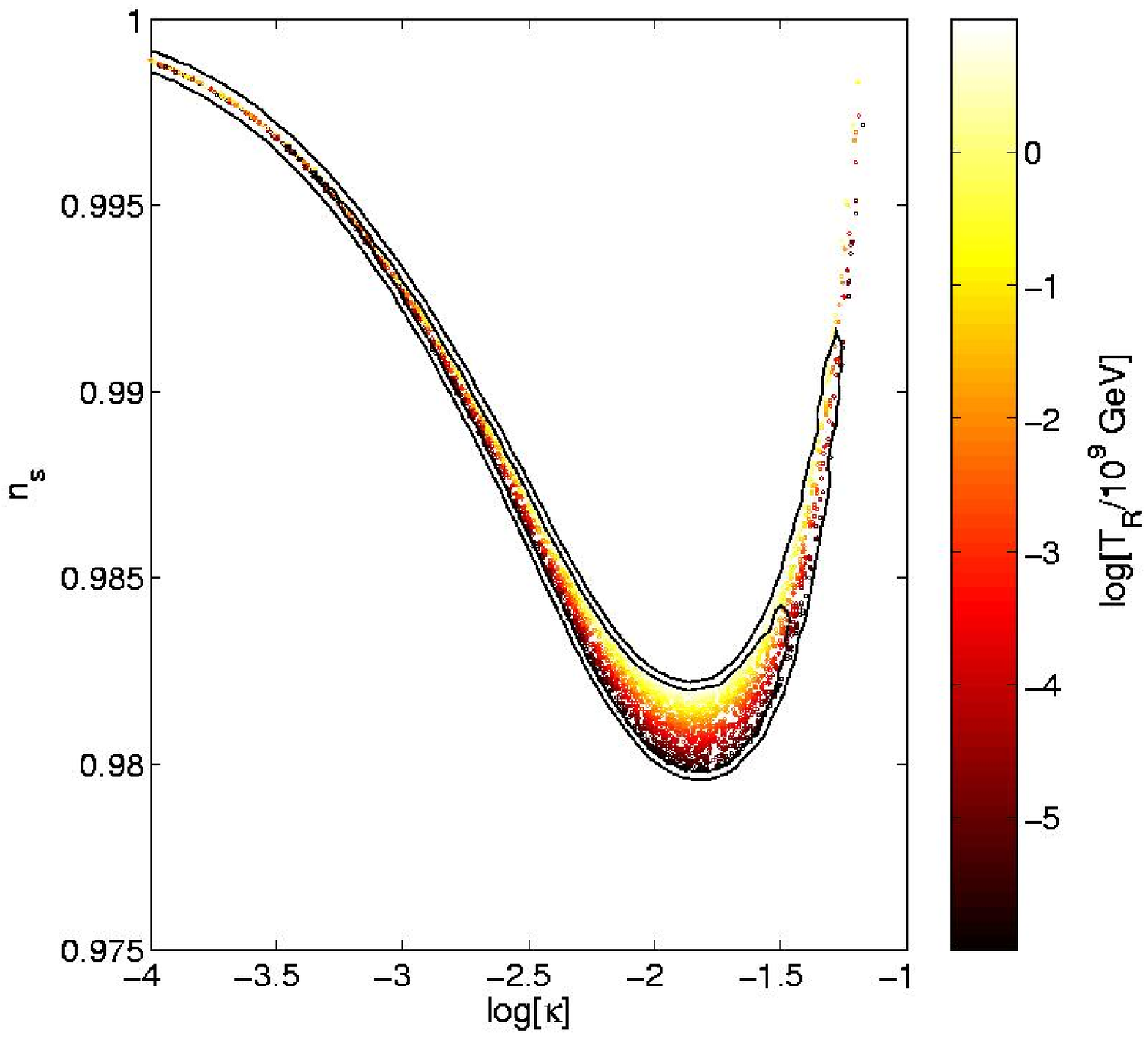,width=7.5cm}
\caption{Constraints on the $F$-term model with strings (left), without strings (right), with minimal SUGRA and $g=0.7$ but allowing $T_{\rm R}$ to vary between $1{\rm TeV}$ and $10^{10}{\rm GeV}$.
The coloured dots encode the reheat temperature $T_{\rm R}$.
\label{figure:FTERMres2}}
\end{figure}

We have also performed an analysis with $T_{\rm R}$ allowed to vary between $1\,{\rm TeV}$ and $10^{10}{\rm GeV}$. The results are not changed significantly, which is what should be expected; $T_{\rm R}$ only modifies $N_{\rm e}$ logarithmically, and the important quantities, $n_{s}$ and $P_{\cal R}$, are only changed by a few percent. The effect of varying $T_{\rm R}$ is illustrated in Fig.~\ref{figure:FTERMres2}.

\begin{figure}[htbp]
\epsfig{file=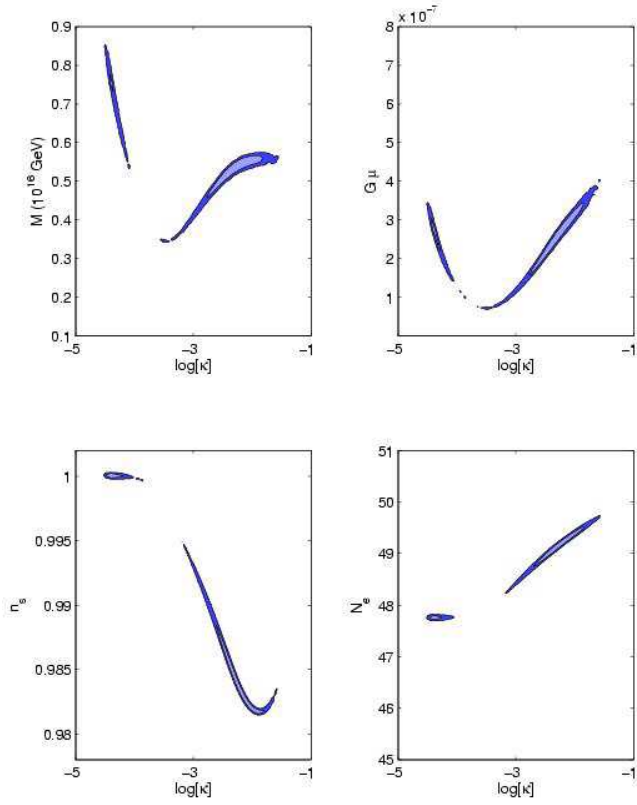,width=8.5cm}
\caption{Constraints on the $F$-term model with minimal SUGRA, curvature and tadpole corrections with $a_S=1{\rm TeV}$, $g=0.7$, $T_{\rm R} =10^9 {\rm GeV}$ and strings.
\label{figure:FTERMres3}}
\end{figure}

In addition, we have performed an analysis where the tadpole
(with $a_S= 1\, {\rm TeV}$) and curvature (with $Q=m_{\rm Pl}/\sqrt{8\pi}$)
corrections to the potential are included. These results are illustrated in Fig.~\ref{figure:FTERMres3} and show a bimodal likelihood surface in the $\kappa-M$ plane. There is an additional allowed region for small $\kappa$, which is absent in the case
without the tadpole contribution. This is due to the fact, that for small
$\kappa$, the parameter $M$ and, hence, also the string tension increase again,
allowing for $n_{s}$ to be close to one.

Constraints on the model parameters for hybrid inflation have been investigated
previously by a number of authors. The most straightforward approach is to
calculate the value of $M$ for a given $\kappa$ in accordance with the
observed value for $P_{\cal R}$. The string contribution to the temperature
fluctuations is proportional to the string tension, which is by~(\ref{nonBogol}) a function of $\kappa$ and $M$. Imposing the reported results (for example, ref.~\cite{pog}) for an upper bound on the string contribution therefore allows one to constrain the parameter $\kappa$. The above approach is valid, since
$P_{\cal R}$ is tightly constrained, and for a given $\kappa$,
$M$ may only vary in a small range.  Rocher and Sakellariadou~\cite{RoSa} obtain $\kappa \stackrel{<}{{}_\sim} 9 \times 10^{-5}$ as an upper bound. However, Jeannerot and Postma~\cite{JP1,JP2} point out that taking
account of the corrections~(\ref{nonBogol}) due to deviations from the Bogomol'nyi limit leads to the significant relaxation to $\kappa \stackrel{<}{{}_\sim} 5 \times 10^{-2}$, under the assumption that strings
contribute less than 10\% to the power spectrum at $\ell=4$; a result
which is in accordance with our analysis.
These approaches do not render a precision determination of the allowed
range for the parameter $\kappa$ due
to the uncertainty in the upper bound on the string contribution, a shortcoming
which we resolve here. More important, unlike the approach of refs.~\cite{RoSa,JP1,JP2}, the methods presented here fully address the impact of the presence of the string network on the preferred values of the other cosmological parameters, in particular $n_{s}$. We note that~\cite{RoSa,JP1,JP2} appeared
before the WMAP3 data, which makes high precision constraints on $n_{s}$
possible, became available.

We also note the work by Fraisse~\cite{fraisse05,fraisse06},
who has performed a fit for the standard six parameters and the relative contribution of topological defects as the seventh, comparable to our analysis in
section~\ref{sec:blue}. While the paper based on the WMAP1 data~\cite{fraisse05} contains plots indicating qualitatively the same dependence of $\Omega_{\rm b}$ and $n_{s}$ on the defect contribution as we present in Fig.~\ref{fig:7param}, such a presentation is not given in the discussion of the
analysis of the WMAP3 data~\cite{fraisse06} where it is more relevant.
Fraisse's study does not, however, exploit the relation between $\kappa$ and the spectral index $n_{s}$ and therefore comes short of giving a precision determination of the parameter $\kappa$. Moreover, the deviation from the Bogomol'nyi limit is not taken into account, leading to a far too tight constraint on $\kappa$.

\subsection{$D$-term models}
\label{sec:dterm_res}

\begin{figure}[htbp]
\epsfig{file=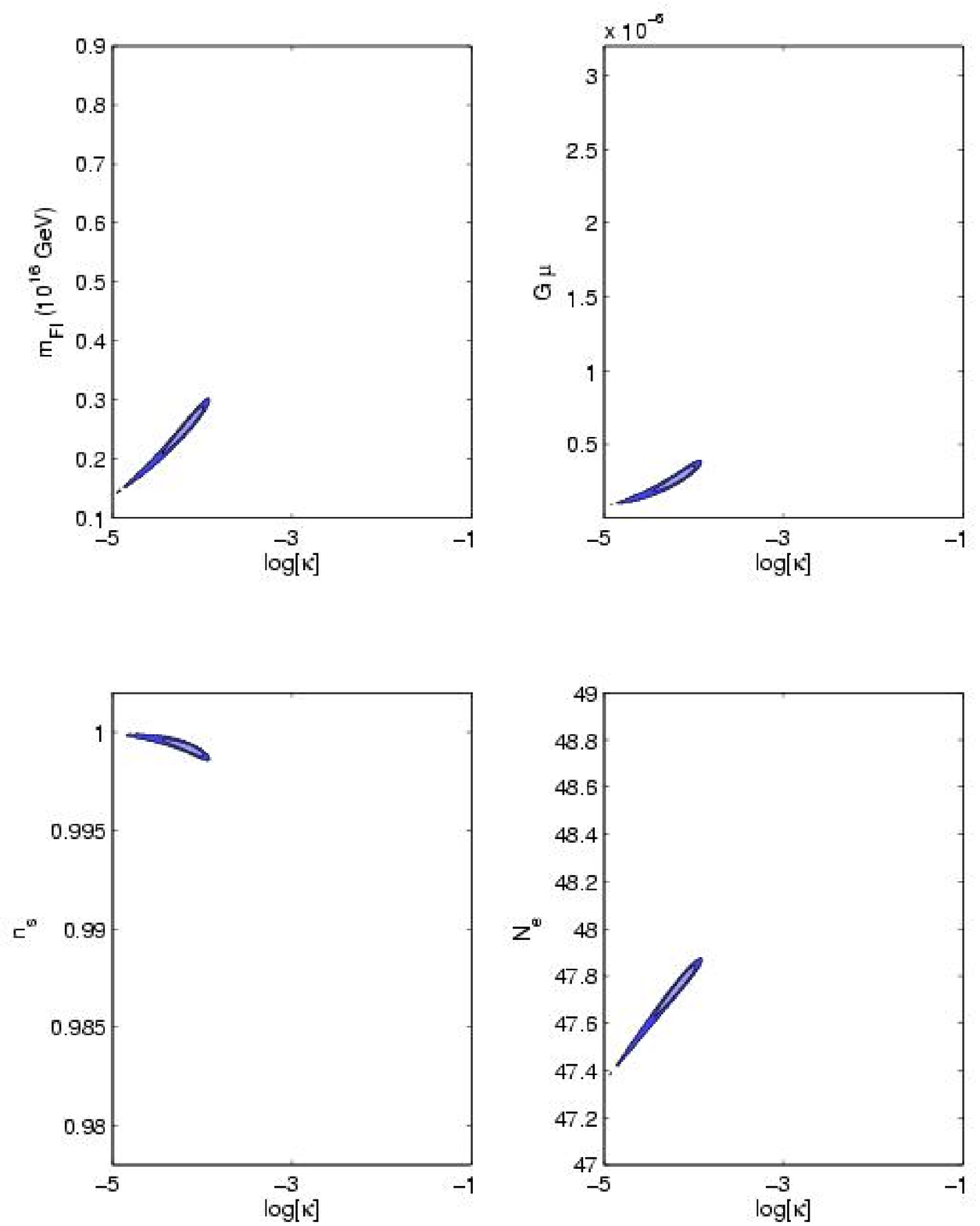,width=8.5cm}
\epsfig{file=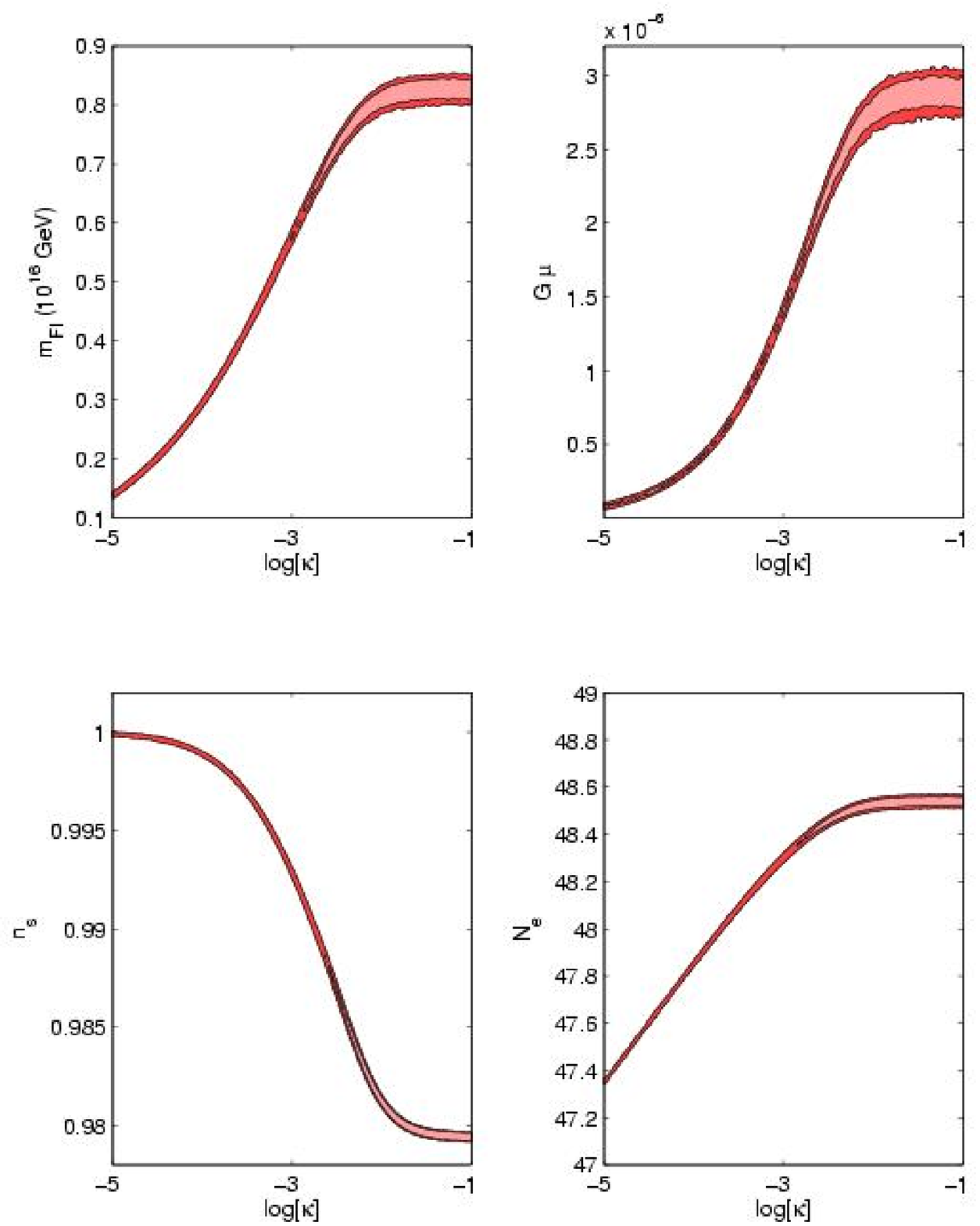,width=8.5cm}
\caption{Constraints on the $D$-term model with $g=10^{-3}$ and  $T_{\rm R} =10^9 {\rm GeV}$. On the left are the allowed regions when strings are included and on the right when they are not included. Note the very different scale for $G\mu$ with and without strings.}
\label{figure:DTERMres}
\end{figure}

We have also investigated constraints on $\kappa$ and $m_{\rm FI}$ for $D$-term inflation models, which are quantitatively different from those for $F$-term inflation. We initially  fix $g=10^{-3}$ and $T_{\rm R}=10^9{\rm GeV}$. The results including the string component and, for comparison, artificially excluding it are presented in Fig.~\ref{figure:DTERMres}. 

For the scenario with strings, we find for the parameters
$\log\kappa= -4.24 \pm 0.19 $, $m_{\rm FI}= (0.24 \pm 0.03)\,{\rm GeV} $
and the likelihood $-2\log{\cal L}=11305.0$ for the best fit model. 
When compared to the $F$-term model, significantly lower values for $\kappa$
are preferred, which in turn also induces lower preferred values for
$m_{\rm FI}$ in comparison to $M$. The reason for this can be seen when the strings are not included -- a wide range of values for $\kappa$ are allowed by the data, but those with $\kappa>10^{-4}$ correspond to string tensions which are excluded. Since the strings satisfy the Bogomol'nyi limit in this case, the string bound is much more restrictive than in the $F$-term model.
When the string contribution to the power spectrum is ignored, we find for the likelihood function of the best-fit model $-2\log{\cal L}=11307.9$.

\begin{figure}[htbp]
\epsfig{file=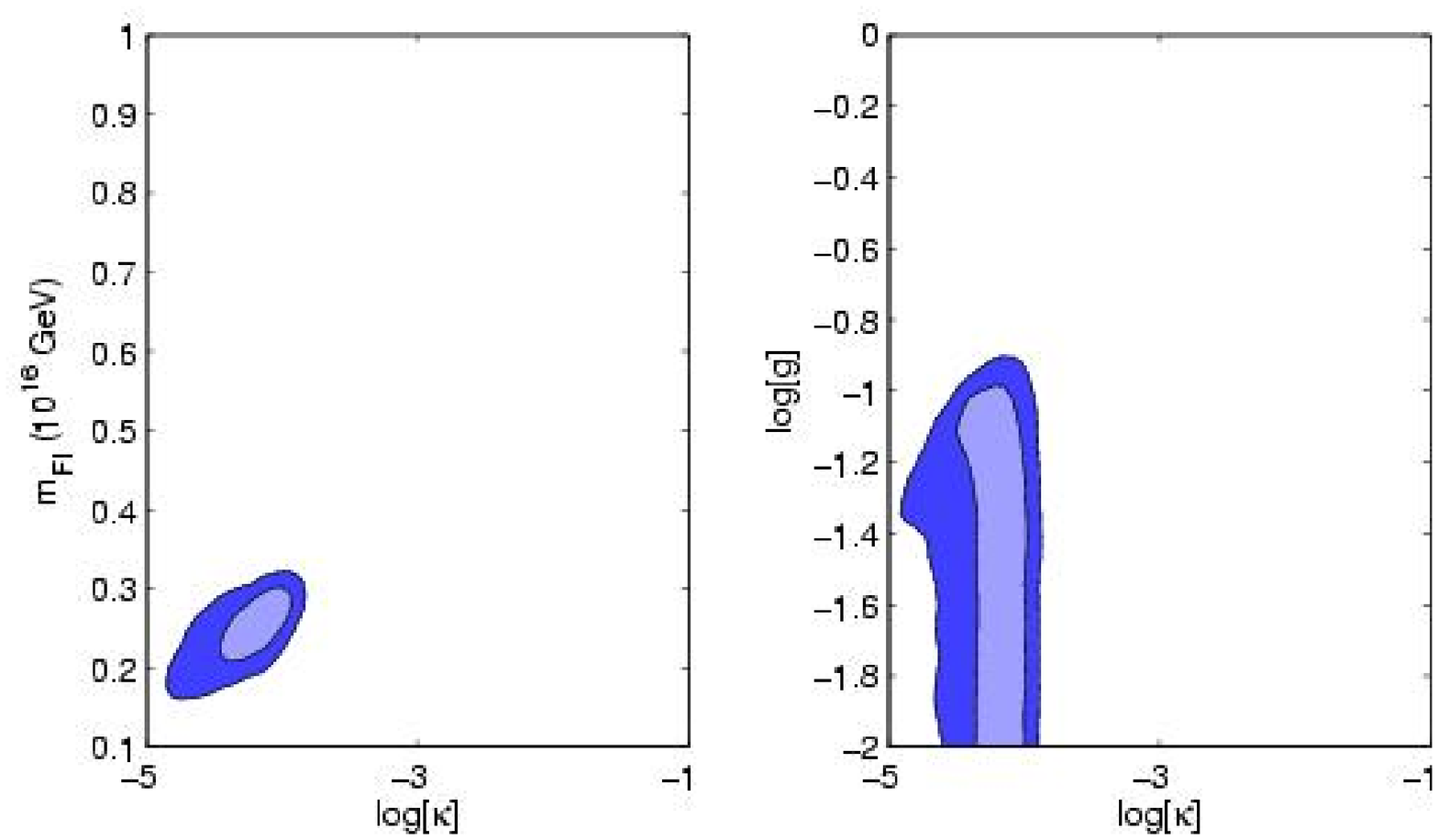,width=8.5cm}
\epsfig{file=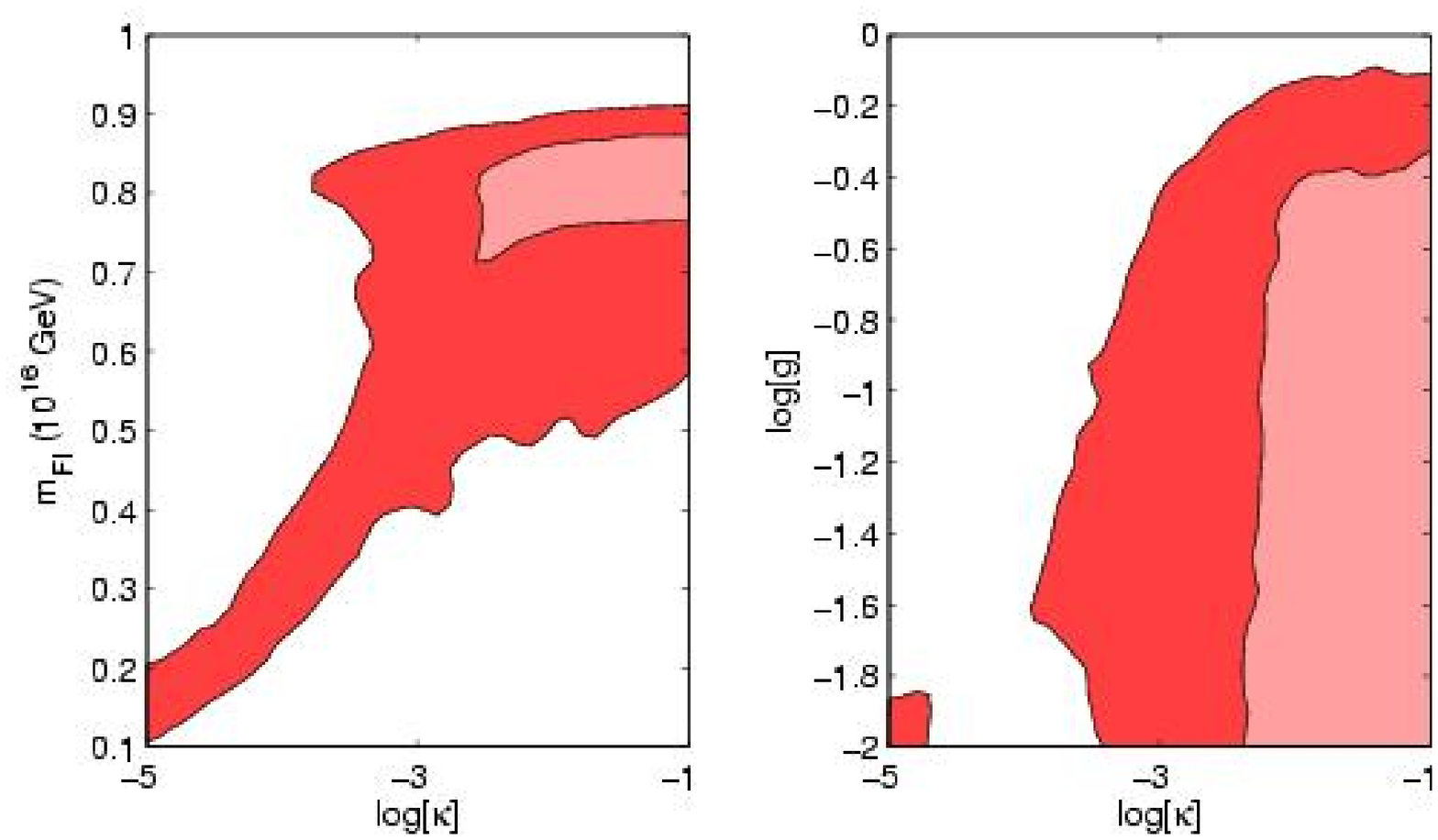,width=8.5cm}
\caption{Constraints on the $D$-term model with $T_{\rm R} =10^9 {\rm GeV}$ and variable $g$. On the left are the allowed regions when strings are included and on the right when they are not included. }
\label{figure:DTERMres2}
\end{figure}

Following our discussion in section~\ref{Dterm}, the results for $g=10^{-3}$
also apply for all values $g \stackrel{<}{{}_\sim} 0.1$. This degeneracy can
be seen from Fig~\ref{figure:dterm2}. For larger values of $g$, the degeneracy is broken, but these points in parameter space are ruled out due to the violation of slow-roll conditions. The $2-\sigma$ upper bound on $g<0.092$ when strings are included and $g<0.44$ when strings are not included; see
Fig.~\ref{figure:DTERMres2}.

\subsection{$F$-term inflation with non-minimal SUGRA}
\label{sec:nonmin}

\begin{figure}[htbp]
\epsfig{file=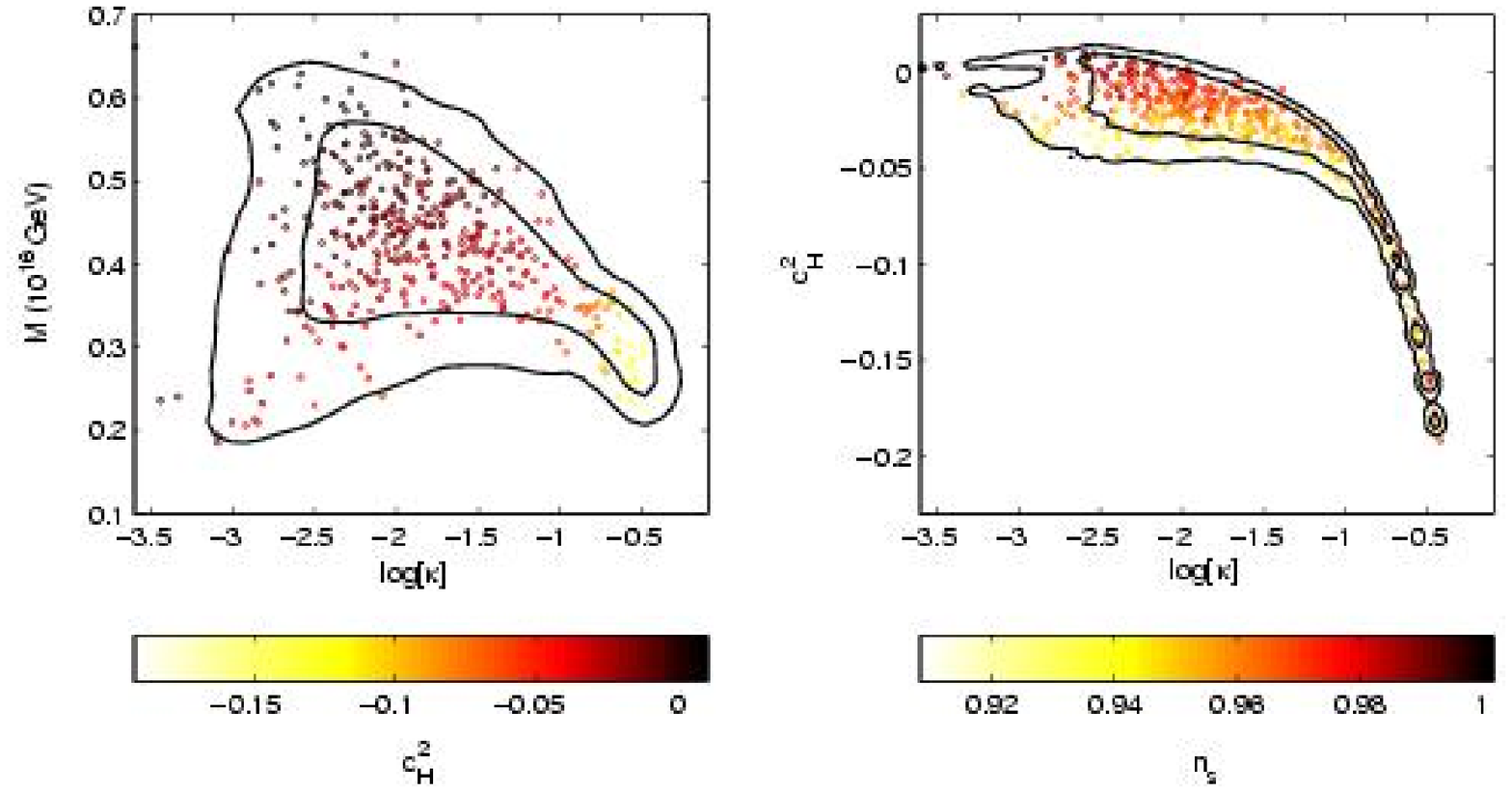,width=10.0cm}
\caption{Constraints on the $F$-term model with strings,
non-minimal SUGRA, tadpole and curvature corrections,
$a_S=1{\rm TeV}$, $g=0.7$ and $T_{\rm R} =10^9 {\rm GeV}$.
\label{figure:FTERMCH2str}}
\end{figure}

\begin{figure}[htbp]
\epsfig{file=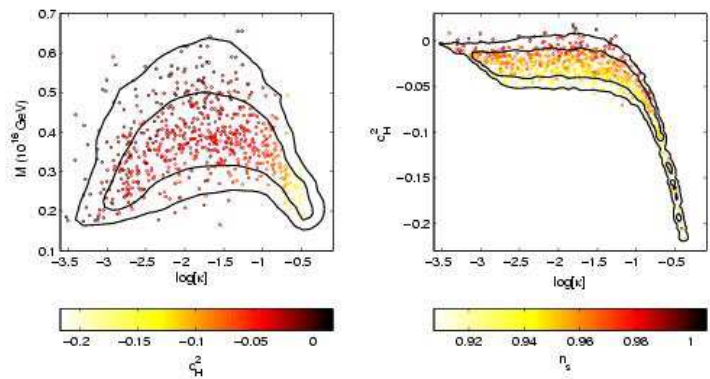,width=10.0cm}
\caption{Constraints on the $F$-term model without including strings, with
non-minimal SUGRA, tadpole and curvature corrections, $a_S=1{\rm TeV}$, 
$g=0.7$ and $T_{\rm R} =10^9 {\rm GeV}$.
\label{figure:FTERMCH2nonstr}}
\end{figure}

Here, we study the $F$-term model with a non-zero parameter $c_H^2$
for the SUGRA contribution~(\ref{V:SUGRA}) and we treat $c_{H}^{2}$ as an additional free parameter in the MCMC analysis.
The results as presented in Figs.~\ref{figure:FTERMCH2str}
and~\ref{figure:FTERMCH2nonstr} show that
negative values for $c_H^2$ lead to an enhancement of the red-tilt of
the spectral index, which is why this parameter originally was
considered~\cite{BasShaKi}. When strings are included,
it can be seen from Fig.~\ref{figure:FTERMCH2str}
that there is a region for the parameter $\kappa$ where a non-zero $c_H^2$
is not required. We find the best fit values for the parameters
$\log \kappa=-1.87 \pm 0.66$, $M=(0.417 \pm 0.093)\times 10^{16}{\rm GeV}$,
$c_H^2=-0.030\pm0.035$
and the overall likelihood $-2\log {\cal L}=11302.6$.
For comparison, we also show in
Fig.~\ref{figure:FTERMCH2nonstr} the results for
the model without including strings,
where indeed negative values for $c_H^2$ are
favoured over the allowed range for $\kappa$. It is worth pointing out
that the allowed range for $\kappa$ reaches up to values
$\kappa\approx 0.4$, whereas in the models with $c_H^2=0$, we find
$\kappa \stackrel{<}{{}_\sim} 0.08$. This is because the minimal SUGRA
correction, which induces a blue tilt of the spectral index, can to some
extent be balanced by negative values for $c_H^2$. However, for such large
values of $\kappa \stackrel{>}{{}_\sim} 0.1$, $c_H^2$ has to be tuned rather
strongly. The poor convergence of the Markov Chains in that region is apparent
by inspecting Figs.~\ref{figure:FTERMCH2str} and~\ref{figure:FTERMCH2nonstr}.


\section{Discussion and conclusions}

We have computed up-to-date precision constraints on the parameters $\kappa$ and $M$ for $F$-term inflation, and $\kappa$ and $m_{\rm FI}$ for $D$-term inflation. Assuming that the data is correct and that one of these minimal SUSY hybrid inflation models is correct, we have shown that one can measure accurately parameters of supersymmetric models at Grand Unified scales. In particular, we have shown that the inclusion of the effects of strings is crucial to establish correct constraints. For comparison, we have performed most analyses also for the case where the string contribution is artificially excluded. Throughout our discussion we have also highlighted potential corrections to the constraints for the most minimal models, in particular we investigated the effect of the leading non-minimal SUGRA correction and of the tadpole induced by soft SUSY breaking.

Without taking account of strings, one may infer from the WMAP3
data~\cite{wmap} that the minimal $F$ term models, which predict
$n_{s} \approx 0.98$, are in tension with the data at
$2-\sigma$ level and that the small coupling domain of hybrid
inflation, where $n_{s} \approx 1$ is ruled out at
$3-\sigma$ level~\cite{PrecCon,JP3}. The latter constraint would in
fact rule out $D$-term inflation, which requires the parameter
$\kappa$ to be small in order not to have an excess contribution of
strings to the perturbation spectrum. However, the self-consistent
analysis performed here reveals that $D$-term inflation is not yet
ruled out and, moreover, that the minimal six parameter $F$-term
models with strings fit the data as well as the standard six parameter
model.

There are also some theoretical modelling uncertainties associated with establishing the cosmic string spectrum. We have shown that if $1.3\le\beta_r\le 2.8$ then the effect on the string power spectrum is around $20\%$, which is $\sim 1\%$ on the total. Sensible variations of $\xi_{r}$ and $\langle v^2\rangle_{r}^{1/2}$ also induce variations of $<20\%$. It appears that further refinement of the string power spectrum beyond this level of understanding is unnecessary for obtaining accurate constraints on hybrid inflation models.

In addition to theoretical uncertainties there are a number of other issues associated with the use of the data which need to be carefully assessed: the discrepancy between the data and a SUSY hybrid inflation model with no strings and $n_{s}\approx 0.98$ is only around $2-\sigma$ (95\% confidence level). Chief amongst the uncertainties is how the polarized foregrounds are extracted and how line-of-sight effects such as gravitational lensing~\cite{CL} and the Sunyaev-Zeldovich (SZ) effect are dealt with. For example in the analysis performed by the WMAP team an SZ contribution was included in the analysis with an amplitude which was marginalized over. Since this contribution to the power spectrum will be increasing with $\ell$, such an analysis will tend to lower the best fit value of $n_{s}$. Conversely, taking into account the gravitational lensing effect will have a tendency to increase the value of $n_{s}$. We have included neither in our analysis, presuming that they will cancel each other out. Once even higher precision data is available, from for example the PLANCK satellite, these effects may come to dominate the systematics.

\begin{figure}[htbp]
\epsfig{file=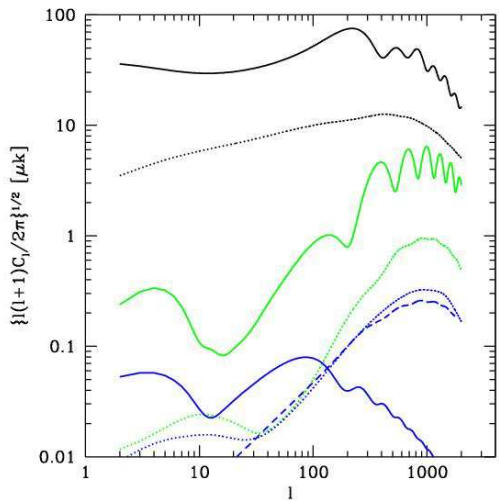,width=7.0cm}
\caption{Components of the temperature and polarization power spectra for the best-fitting $F$-term inflation model with an artificially large value of $r=0.1$ at $k=0.05{\rm Mpc}^{-1}$. The solid lines are the adiabatic component, the dotted lines the string component and the dashed line is the $B$-mode component due to the gravitational lensing of the $E$-mode polarization of the adiabatic component. For the adiabatic and string components, the three curves (top to bottom) are the temperature power spectrum and that for the $E$- and $B$-mode polarization.}
\label{figure:bmode}
\end{figure}

There is an important observational signature of these models which may be in reach in the near future. Inflationary models with non-zero $r$ predict that there will be $B$-mode polarization on large scales peaking around $\ell\approx 100$, but as we have already noted the SUSY hybrid inflation models predict $r<10^{-4}$ and a signal this weak is unlikely to ever be detected. However, as we have already pointed out the anisotropies created by cosmic strings create $B$-mode polarization since they do not distinguish between scalar, vector and tensor anisotropies. Fig.~\ref{figure:bmode} shows both the adiabatic and cosmic string components to the temperature anisotropies and polarization for a model with an artificially large value of $r=0.1$ at $k=0.05{\rm Mpc}^{-1}$. It can be seen that there is a $B$-mode polarization signal due to the cosmic string component, which has an amplitude of $\approx 0.3\mu{\rm K}$ at around $\ell\approx 1000$. This has very different characteristics to that due to adiabatic tensor perturbations and is of similar amplitude to the contribution expected due to the conversion of $E$-mode polarization into $B$-mode by gravitational lensing. In ref.~\cite{Seljak:2006hi} it was pointed out that string tensions of as low as $G\mu\approx 10^{-9}$ might be detectable in future CMB polarization missions if one is able to ``clean'' the lensing contribution using high resolution observations.

\begin{figure}[htbp]
\epsfig{file=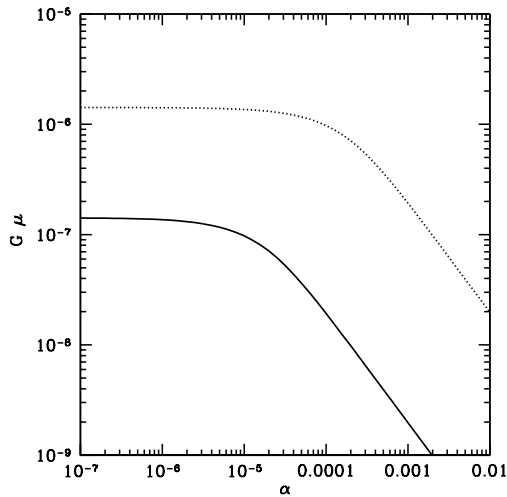,width=7.0cm}
\caption{Constraints on $G\mu$ from due to the absence of timing noise in pulsars against the loop production size $\alpha$. The region above the solid line is ruled out at $2-\sigma$ if $\Omega_{\rm g}h^2<2\times 10^{-9}$ for $f\approx 2\times 10^{-9}{\rm Hz}$. See note added in proof for the explanation of the dashed line.}
\label{figure:pulsar}
\end{figure}

Finally we should point out that a network of cosmic strings with $G\mu\sim 10^{-7}$ will lead to a number of other potentially observable effects. In particular, the decay of cosmic string loops could create  a stochastic background of gravitational waves (see ref.~\cite{Caldwell:1996en} and references therein) if the dominant decay channel for the strings is gravitation. Such a background is constrained by the lack of timing residuals in the observations of pulsars; the most recent constraint being that $\Omega_{\rm g}h^2<2\times 10^{-9}$ at frequencies of $f\approx 2\times 10^{-9}{\rm Hz}$~\cite{pulsar}. If we assume that the absolute lower bound on the string spectrum is given by the ``red-noise'' spectrum generated by the decay of string loops in the radiation era, then one can use various parameters measured by string network simulations~\cite{BB,AS} and the measured value of $\Omega_{\rm m}$ to compute a bound on $G\mu$ as a function of the loop production size relative to the horizon, $\alpha$. The results of doing this are presented in Fig.~\ref{figure:pulsar}. We see that for small $\alpha$ there is a plateau with $G\mu<1.4\times 10^{-7}$ and for larger values of $\alpha$ (which are probably less likely) more tight constraints are possible. These results are considerably more stringent than those presented in ref.~\cite{Caldwell:1996en} since at that time the limit $\Omega_{\rm g}h^2<9\times 10^{-8}$ was used.

Taken at face value these results appear to further constrain, but do not yet rule out, the $F$-term scenarios under consideration here to a narrow range of $\log\kappa\approx -3$ and $M\approx 4\times 10^{15}{\rm GeV}$, with a qualitatively similar situation in the $D$-term case. There are, however, numerous uncertainties, particularly in the details of string evolution which could substantially change the conclusions and therefore at this stage we feel that it would not be sensible to include these observations in our likelihood analysis. We have already noted that the anisotropy power spectrum that we would observe in the CMB is not that sensitive to the details of string evolution since it is sub-dominant and therefore it is unlikely that one would be able to unequivocally rule in or out these models on the basis of CMB measurements. Hence, it appears that improved observations of pulsar timing and a significantly better understanding of string network evolution, directed towards the pulsar bound, would be the best way of constraining these scenarios further.   
 
\section*{Acknowledgements}
We are grateful to Mark Hindmarsh and Apostolos Pilaftsis for helpful comments.

\section*{Note added in proof}
Ref.~\cite{Jenet} appeared recently claiming that in fact the pulsar constraint is $\Omega_{\rm g}h^2 < 2 \times 10^{-8}$. If this is the case, the constraint on $G\mu$ is reduced by a factor of 10 as illustrated by the dashed line in Fig.~\ref{figure:pulsar}.

\appendix*
\section{Summary of results\label{App:Summary}}

We summarize here the best-fit parameters together with the $1-\sigma$
confidence intervals for the various $F$-term scenarios in
table~\ref{FTermSummary} and for the $D$-term model in
table~\ref{DTermSummary}. For comparison, we have included in each table
the results for the standard six parameter fit and for the seven parameter
fit including strings.

\begin{center}
\begin{sidewaystable}

\begin{tabular}{|c||c|c||c|c||c|c||c||c|c||} \hline
& \multicolumn{9}{|c|}{Model} \\ \hline
 Parameter &  I(s) &  I(ns) & II (s) & II(ns) &  III(s) & III(ns) & IV (s)  &  V(s)  & V(ns)  \\ \hline

$\Omega_{\rm b}$  & $0.0237 \pm 0.0015$ & $0.0225 \pm 0.0007$ & $0.0255 \pm 0.0009$ & $0.0235 \pm 0.0005$ & $0.0256 \pm 0.0009$ & $0.0234 \pm 0.0004$ & $0.0257 \pm 0.0010$ & $0.0242 \pm 0.0014$ & $0.0225 \pm 0.0007$ \\ \hline

$\Omega_{\rm c}$  & $0.103 \pm 0.007$ & $0.106 \pm 0.007$ & $0.100 \pm 0.007$ & $0.102 \pm 0.007$ & $0.100 \pm 0.007$ & $0.101 \pm 0.006$ & $0.100 \pm 0.007$ &  $0.103 \pm 0.007$ & $0.105 \pm 0.007$ \\ \hline

$\theta_{\rm A}$   & $1.044 \pm 0.004$ & $1.042 \pm 0.003$ & $1.048 \pm 0.003$ & $1.045 \pm 0.003$ & $1.048 \pm 0.003$ & $1.044 \pm 0.002$ & $1.049 \pm 0.003$ & $1.045 \pm 0.004$ & $1.042 \pm 0.003$ \\ \hline

$\tau_{\rm R}$   & $0.097 \pm 0.032$ & $0.092 \pm 0.029$ & $0.113 \pm 0.029$ & $0.123 \pm 0.028$ & $0.112 \pm 0.026$ & $0.123 \pm 0.024$ & $0.115 \pm 0.029$ & $0.095 \pm 0.030$ & $0.090 \pm 0.030$ \\ \hline

$\log (10^{10} P_{\cal R})$   & $3.00 \pm 0.07$ & $3.03 \pm 0.06$ & $3.00 \pm 0.07$ & $3.09 \pm 0.05$  & $3.00 \pm 0.06$ & $3.09 \pm 0.05$ & $3.00 \pm 0.07$ &$2.98 \pm 0.06$ & $3.02 \pm 0.06$ \\ \hline \hline

$n_{s} $  & $0.964 \pm 0.019$ & $0.956 \pm 0.016$ & $0.985 \pm 0.004$  & $0.987 \pm 0.006$ & $0.984 \pm 0.003$ & $0.986 \pm 0.006$ & $0.987 \pm 0.006$ & $0.964 \pm 0.016$ & $0.956 \pm 0.016$ \\ \hline

$\log \kappa$  & - & - & $-2.34 \pm 0.38$ & $-2.40 \pm 0.88$ & $-2.32 \pm 0.35$ & $-2.28 \pm 0.78$ & $-2.58 \pm 0.76$ & $-1.87 \pm 0.66$ & $-1.82 \pm 0.72$ \\ \hline

$\log(T_{\rm R} /10^{9} {\rm GeV})$   & - & - & 0.0 & 0.0 & $-2.5 \pm 2.0$ & $-2.6 \pm 2.0$ & 0.0 & 0.0 & 0.0  \\ \hline

$M /10^{16} {\rm GeV}$  & - & - & $0.518 \pm 0.059$ & $0.495 \pm 0.139$ & $0.525 \pm 0.054$ & $0.518 \pm 0.122$ & $0.549 \pm 0.079$ & $0.417 \pm 0.093$ & $0.373 \pm 0.102$ \\ \hline

$c_H^{2}$  & - & - & - & - & - & - & - & $-0.030 \pm 0.035 $ & $-0.038 \pm 0.038 $ \\ \hline

$G \mu /10^{-7}$ & $ <3.0 $ & - & $2.50 \pm 0.65$ & $2.81 \pm 1.61$ & $2.56 \pm 0.61$ & $3.02 \pm 1.51$ & $2.54 \pm 0.60$ & $2.07 \pm 0.72$ & $1.85 \pm 1.12$  \\ \hline

$N_{\rm e}$  & - & - & $49.1 \pm 0.5$ & $49.0 \pm 0.9$ & $47.2 \pm 1.6$ & $47.1 \pm 1.7$  & $48.9 \pm 0.5$ & $49.3 \pm 0.5$ & $49.2 \pm 0.6$ \\ \hline

$\Omega_{\rm m} $  & $0.217 \pm 0.033$ & $0.237 \pm 0.031$ & $0.191 \pm 0.023$ & $0.211 \pm 0.026$ & $0.191 \pm 0.023$ & $0.209 \pm 0.025$ & $0.190 \pm 0.024$ & $0.214 \pm 0.031$ &  $0.236 \pm 0.030$ \\ \hline

$\Omega_{\rm \Lambda} $   & $0.783 \pm 0.033$ & $0.763 \pm 0.031$ & $0.809 \pm 0.023$ & $0.789 \pm 0.026$ & $0.809 \pm 0.023$ & $0.791 \pm 0.025$ & $0.810 \pm 0.024$ & $0.786 \pm 0.031$ & $0.764 \pm 0.030$ \\ \hline

$t_{0}/{\rm Gyr}$   & $13.44 \pm 0.26$ & $13.66 \pm 0.15 $ & $13.12 \pm 0.15$ & $13.42 \pm 0.10$ & $13.11 \pm 0.15$ & $13.45 \pm 0.09$ & $13.10 \pm 0.16$ &  $13.37 \pm 0.25$ & $13.66 \pm 0.15$ \\ \hline

$z_{\rm re}  $   & $11.2 \pm 2.5$ & $11.3 \pm 2.5$ & $12.0 \pm 2.2$ & $13.4 \pm 2.1$ & $11.9 \pm 1.9$ & $13.4 \pm 1.8$ & $12.0 \pm 2.2$ & $10.9 \pm 2.3$ & $11.1 \pm 2.5$ \\ \hline

$h  $   & $0.77 \pm 0.04$ & $0.74 \pm 0.03$ & $0.81 \pm 0.03$ & $0.77 \pm 0.03$ & $0.81 \pm 0.03$ & $0.77 \pm 0.03$ & $0.82 \pm 0.03$ & $0.78 \pm 0.04$ & $0.74 \pm 0.03$ \\ \hline \hline

$-2 \log \mathcal{L}$ & 11302.8 & 11305.5 & 11303.3 & 11308.4 & 11303.2 & 11308.0 & 11303.2 & 11302.6 &  11305.5 \\ \hline \hline

\end{tabular}

\caption{ \label{FTermSummary}
Constraints on $F$-term inflationary scenarios. In each case we quote results with the string contribution (s) and without the string contribution to the CMB (ns). Model I is the basic fit, with no inflationary input parameters, and is a six parameter (in the case of no strings) and seven parameter (no strings) fit. Model II is the basic minimal $F$-term model, with six input parameters, while model III has the reheat temperature of inflation $T_{\rm R}$ as an additional parameter. Model IV includes curvature and tadpole correction to the potential (and is also a six parameter fit) and model V includes the non-minimal SUGRA parameter $c_H^{2}$ as an additional parameter.}
\end{sidewaystable}
\end{center}

\begin{center}
\begin{sidewaystable}
\begin{tabular}{|c||c|c||c|c||c|c||} \hline
& \multicolumn{6}{|c|}{Model} \\ \hline
 Parameter &  I(s) &  I(ns) & VI (s) & VI(ns) &  VII(s) & VII(ns)  \\ \hline

$\Omega_{\rm b}$  & $0.0237 \pm 0.0015$ & $0.0225 \pm 0.0007$ &  $0.0261 \pm 0.0011$ & $0.0234 \pm 0.0005$ & $0.0263 \pm 0.0011$ & $0.0234 \pm 0.0004$ \\ \hline

$\Omega_{\rm c}$  & $0.103 \pm 0.007$ & $0.106 \pm 0.007$ & $0.098 \pm 0.007$ & $0.102 \pm 0.007$ & $0.098 \pm 0.007$ & $0.102 \pm 0.007$ \\ \hline

$\theta_{\rm A}$   & $1.044 \pm 0.004$ & $1.042 \pm 0.003$ & $1.050 \pm 0.003$  & $1.045 \pm 0.003$ & $1.050 \pm 0.003$ & $1.045 \pm 0.003$ \\ \hline

$\tau_{\rm R}$   & $0.097 \pm 0.032$ & $0.092 \pm 0.029$ & $0.131 \pm 0.028$ & $0.119 \pm 0.029$ & $0.131 \pm 0.029$ & $0.118 \pm 0.028$ \\ \hline

$\log (10^{10} P_{\cal R})$   & $3.00 \pm 0.07$ & $3.03 \pm 0.06$  & $3.04 \pm 0.07$ & $3.08 \pm 0.06$ & $3.03 \pm 0.07$ & $3.08 \pm 0.05$ \\ \hline \hline

$n_{s} $  & $0.964 \pm 0.019$ & $0.956 \pm 0.016$ & $0.9993 \pm 0.0003$  & $0.984 \pm 0.006$ & $1.000 \pm 0.001$ & $0.985 \pm 0.006$ \\ \hline

$\log \kappa$  & - & - & $-4.24 \pm 0.19$ & $-2.10 \pm 0.89$ & $-4.24 \pm 0.20$ & $-1.97 \pm 0.76$ \\ \hline

$\log(T_{\rm R} /10^{9} {\rm GeV})$   & - & -  & 0.0 & 0.0 & 0.0 & 0.0 \\ \hline

$m_{\rm FI} /10^{16} {\rm GeV}$  & - & - & $0.245 \pm 0.031$ & $0.730 \pm 0.171$ & $0.249 \pm 0.029$ & $0.754 \pm 0.135$ \\ \hline

$\log(g)$  & - & - & -3.0 & -3.0  & $<-1.0$ & $<-0.36$\\ \hline

$G \mu /10^{-7}$ & $ <3.0 $ & - & $2.56 \pm 0.61$ &  $23.7 \pm 8.1$ & $2.65 \pm 0.57$ &  $24.7 \pm 6.7$ \\ \hline

$N_{\rm e}$  & - & - & $47.7 \pm 0.1$ & $48.4 \pm 0.3$ & $48.9 \pm 0.3$ & $49.9 \pm 0.4$ \\ \hline

$\Omega_{\rm m} $  & $0.217 \pm 0.033$ & $0.237 \pm 0.031$ &  $0.181 \pm 0.022$ & $0.213 \pm 0.026$ & $0.179 \pm 0.022$ & $0.213 \pm 0.026$ \\ \hline

$\Omega_{\rm \Lambda} $   & $0.783 \pm 0.033$ & $0.763 \pm 0.031$ & $0.819 \pm 0.022$  & $0.787 \pm 0.026$ & $0.821 \pm 0.022$ & $0.787 \pm 0.026$ \\ \hline

$t_{0}/{\rm Gyr}$   & $13.44 \pm 0.26$ & $13.66 \pm 0.15 $ & $13.00 \pm 0.17$  & $13.45 \pm 0.10$ & $12.97 \pm 0.18$ & $13.45 \pm 0.10$ \\ \hline

$z_{\rm re}  $   & $11.2 \pm 2.5$ & $11.3 \pm 2.5$ & $13.0 \pm 2.0$ & $13.1 \pm 2.2$ & $13.0 \pm 2.0$ & $13.1 \pm 2.1$ \\ \hline

$h  $   & $0.77 \pm 0.04$ & $0.74 \pm 0.03$ & $0.83 \pm 0.03$  & $0.77 \pm 0.03$ & $0.84 \pm 0.03$  & $0.77 \pm 0.03$ \\ \hline \hline

$-2 \log \mathcal{L}$ & 11302.8 & 11305.5 & 11305.0 & 11307.9 & 11305.0 & 11308.0 \\ \hline \hline

\end{tabular}
\caption{
\label{DTermSummary}
Constraints on $D$-term inflationary scenarios. In each case we quote results with the string contribution(s) and without the string contribution to the CMB (ns). Model I is the basic fit, with no inflationary input parameters, and is a six parameter (in the case of no strings) and seven parameter (no strings) fit. Model VI is the basic minimal $D$-term model, with six input parameters, while model VII includes the gauge coupling $g$ as an additional parameter, where
we give the $2\sigma$-level upper bounds on $g$.}
\end{sidewaystable}
\end{center}

\end{document}